\documentclass[twocolumn,trackchanges]{aastex62}
\usepackage{amsmath}
\hypersetup{
	colorlinks	= true,
	linkcolor	= red,
	urlcolor	= cyan,
	citecolor	= blue
}

\let\Moon\relax

\let\Jupiter\relax

\let\Libra\relax

\let\Aries\relax

\usepackage{marvosym}
\usepackage{amssymb}

\graphicspath{{./images/}}

\newcommand{\multinest}{\mbox{\textsc{MultiNest} }}
\newcommand{\exorelr}{\mbox{\textsc{ExoReL$^\Re$} }}
\newcommand{\exorel}{\mbox{\textsc{ExoReL} }}
\newcommand{\upsand}{\mbox{$\upsilon$ And e$\ $}}

\turnoffediting

\accepted{February 22, 2020}

\submitjournal{AJ}

\shortauthors{Damiano \& Hu}

\begin{document}
	
	\title{\exorelr: A Bayesian Inverse Retrieval Framework For Exoplanetary Reflected Light Spectra}
	
	\correspondingauthor{Mario Damiano}
	\email{mario.damiano@jpl.nasa.gov}
	
	\author[0000-0002-1830-8260]{Mario Damiano}
	\affiliation{Jet Propulsion Laboratory, California Institute of Technology, Pasadena, CA 91109, USA}
	
	\author[0000-0003-2215-8485]{Renyu Hu}
	\affiliation{Jet Propulsion Laboratory, California Institute of Technology, Pasadena, CA 91109, USA}
	\affiliation{Division of Geological and Planetary Sciences, California Institute of Technology, Pasadena, CA 91125, USA}
	
	\begin{abstract}
		
		
		The high-contrast imaging technique is meant to provide insight into those planets orbiting several astronomical units from their host star. Space missions such as WFIRST, HabEx, and LUVOIR will measure reflected light spectra of cold gaseous and rocky planets. 
		
		To interpret these observations we introduce \exorelr (Exoplanetary Reflected Light Retrieval), a novel Bayesian retrieval framework to retrieve cloud properties and atmospheric structures from exoplanetary reflected light spectra. As a unique feature, it assumes a vertically non-uniform volume mixing ratio profile of water and ammonia, and use it to construct cloud densities. In this way, clouds and molecular mixture ratios are consistent.
		
		We apply \exorelr on three test cases: two exoplanets (\upsand and 47 Uma b) and Jupiter. We show that we are able to retrieve the concentration of methane in the atmosphere, and estimate the position of clouds when the S/N of the spectrum is higher than 15, in line with previous works. Moreover, we described the ability of our model of giving a chemical identity to clouds, and we discussed whether or not we can observe this difference in the planetary reflection spectrum. Finally, we demonstrate how it could be possible to retrieve molecular concentrations (water and ammonia in this work) below the clouds by linking the non-uniform volume mixing ratio profile to the cloud presence. This will help to constrain the concentration of water and ammonia unseen in direct measurements.
		
	\end{abstract}
	
	\keywords{methods: data analysis - methods: statistical - planets and satellites: atmospheres - technique: spectroscopic - radiative transfer}
	
	\section{Introduction} \label{sec:intro}
	
	The diversity observed in the thousands of exoplanets present nowadays in our catalog has extended the horizon of our knowledge of the dynamical, physical, and chemical properties of these alien worlds. This has mostly been made possible by characterizing their atmospheres. Focusing on the gaseous giant planets population, the majority of them are made of hydrogen and helium. Therefore, the relevant questions concern the amounts of all elements other than hydrogen and helium, i.e. the heavy elements, that are present.
	The atmospheres of short-period gaseous planets (these are generally hot or warm), Jupiter- and Neptune-size, have been observed. The emission and transmission spectra have revealed molecular absorption of H$_2$O, CO, CH$_4$, CO$_2$, TiO and VO \citep{Swain2008, Swain2009, Snellen2010, Fraine2014, Evans2016, Sing2016, Damiano2017, Damiano2019, Tsiaras2018} and in some cases the presence of clouds and hazes in the atmosphere \citep{Berta2012, Knutson2014, Sing2016, Barstow2017, Tsiaras2018}. The transit technique has provided most of the current result as it benefits more from target planets being close to their parent stars. However, these planets show a different environment compared to the scenario emerging from the studies conducted in our Solar System planets due to higher irradiation received \citep{Burrows1997, Seager1998, Karkoschka1998}.
	
	The high-contrast imaging technique is poised to provide insight into those planets orbiting several astronomical units from their host star so that their equilibrium temperature is low enough to let different chemical and dynamical behavior emerge (e.g., condensation mechanism, cold trap effects, etc.) with respect to the better studied hot counterparts. This technique has been successfully tested in studying forming star and planet regions \cite{Barman2011, Skemer2014, Macintosh2015}. 
	Future direct-imaging exoplanet space mission and mission concept, e.g. \textit{Wide-Field InfraRed Survey Telescope} (WFIRST, \cite{Spergel2013, Spergel2015}), \textit{Habitable Exoplanet Imaging Mission} (HabEx, \cite{Mennesson2016}), \textit{Large Ultra-Violet/Optical/InfraRed Surveyor} (LUVOIR, \cite{Peterson2017}), and Starshade rendezvous probe\footnote{\url{https://smd-prod.s3.amazonaws.com/science-red/s3fs-public/atoms/files/Starshade2.pdf}}, will have the possibility to observe through high-contrast imaging the starlight reflected by exoplanets, and to unveil their atmospheric structure. 
	Rayleigh scattering, molecular absorption, and scattering and absorption by atmospheric condensates determine the reflection spectra of gaseous exoplanets \citep{Marley1999, Seager2000}. Clouds, if present in the atmosphere, are the primary factor that controls the appearance of an exoplanet. Previous studies have shown that the presence and formation of the clouds are regulated by the atmospheric temperature \citep{Sing2016, Barstow2017}. Assuming an atmospheric elemental abundance the same as the Sun and a suitable atmospheric temperature ($\sim$ 200 - 300 K) gaseous giant exoplanets may have ammonia, water, or silicate clouds in their atmospheres \citep{Sudarsky2000, Sudarsky2003, Burrows2004}. The radiative properties of the clouds are sensitive to the vertical extent and density of the cloudy layers and the sizes of cloud particles \citep{Ackerman2001}. The elemental abundance of the atmosphere also affects the formation of the clouds \citep{Cahoy2010}. For these reasons, reflected light spectra of exoplanets contain rich information on the composition and dynamic processes of the exoplanetary atmosphere. In the wavelength range within $0.4$ and $1.0$ $\mu$m, where the reflection spectroscopy mostly operates, it is possible to probe the molecular signatures of methane, ammonia, and water vapor \citep{Hu2014B2014arXiv1412.7582H, Burrows2014, Marley2014} along with the relative condensates. For example, the Jupiter reflection spectrum (e.g. \cite{Karkoschka1998}) contains different levels of methane absorption which have been used to reject simple models of a single reflective cloud deck, favoring a more complex double-layer cloud structure \citep{Sato1979}.
	
	To interpret a spectrum and extrapolate information from it, a comparison between the observed data and the proposed model should be performed through a statistical inverse modeling. While several transmission and emission spectra inverse retrieval frameworks have been developed and established (e.g. \cite{Irwin2008, Madhusudhan2009, Benneke2012, Waldmann2015B2015ApJ...813...13W, Waldmann2015B2015ApJ...802..107W}), reflected light spectroscopic retrieval models, to date, have just started to be explored. Several models have been proposed (e.g. \cite{Lupu2016, Feng2018, Batalha2019}) but these models use optical properties of clouds (optical depth, scattering albedo, and asymmetry factor) as free model parameters without bounding them to a physical model of the cloud structure (e.g. particle size and chemical cloud identity).
	
	In this work, we present \exorelr (Exoplanetary Reflected Light Retrieval), a novel inverse retrieval framework based on a modified version of \exorel \citep{Hu2019B2019ApJ...887..166H}, which is a cloud formation and radiative transfer model to synthesize the wavelength dependence of the albedo (and therefore planetary flux) of a gaseous planetary atmosphere. \exorelr uses non-constant volume mixing ratio vertical profile of water and ammonia as input to compute the density and the particle size of water and ammonia clouds, as well as a T-P profile consistent with the lapse rate equation.
	This algorithm is used as forward model for the Bayesian sampler \textit{nested sampling} \citep{Skilling2004, Sivia2006, Skilling2006} and its implementation \multinest \citep{Feroz2007, Feroz2009, Feroz2013, Buchner2014} to perform inverse retrieval processes on reflected light spectra. 
	
	
	The parameters adopted in this work are consistent with the gas giant exoplanets scenario that have equivalent orbital distances of 1-6 AU around nearby F-G-K stars. Moreover, we adopted a spectral resolution of R$=$70 for our tests (similar to the WFIRST detector spectral resolution), to explore the reflection spectra of giant exoplanets at 0.4 - 1.0 $\mu$m except for the Jupiter case (Sec. \ref{sec:sol_sys}) that has been studied with a spectral resolution of R$=$120. Results and settings of this work are also generally applicable to future direct imaging mission concepts as they will be sensitive to similar regimes of planetary parameters.
	
	In this paper, we describe our model and the basic concept behind reflection spectroscopy. We will provide insight on the Bayesian analysis and we will present and discuss the results. The manuscript is organized as follow: in Sec. \ref{sec:model} we provide details of \exorelr. In particular, we will discuss the atmospheric structure model, the free parameter space and the details related to the retrieval settings. In Sec. \ref{sec:impact} we will give insight on the impact that each free parameter gives to the albedo spectrum and therefore to the planetary reflected flux. In Sec. \ref{sec:below_cld} and \ref{sec:cld_diff} we will show some of the implications of the set-up adopted in this work. In Sec. \ref{sec:exop} we will explore the performance and the ability of our model to retrieve information from different scenarios by applying it to two exoplanetary test case (\upsand in Sec. \ref{sec:upsand} and 47 Uma b in Sec. \ref{sec:47umab}). In Sec. \ref{sec:sol_sys} we report the results of the analysis of the albedo of Jupiter (Sec. \ref{sec:jup}). In Sec. \ref{sec:discussion} we are going to discuss the results obtained and the implications introduced by this novel model. Finally, in Sec. \ref{sec:conclusion} we will summarize the key points of the paper, and we will discuss the future development of \exorelr.
	
	\section{\exorelr} \label{sec:model}
	\subsection{Amended forward model} \label{sec:structure}
	
	The forward model that synthesizes the planetary geometric albedo and the reflection spectrum is a modified version of the self-consistent \exorel model presented in \cite{Hu2019B2019ApJ...887..166H}. In particular, in \exorel the atmosphere is divided into layers and in each of these, the saturation point of water and ammonia in the gas phase is checked. If one or both reaches the saturation, the humidity is calculated and the relative volume mixing ratio (VMR) vertical profile decreases accordingly. The amount subtracted from the VMR is then used to calculate the physical and optical properties of the clouds.
	
\begin{figure*}[]
		\plotone{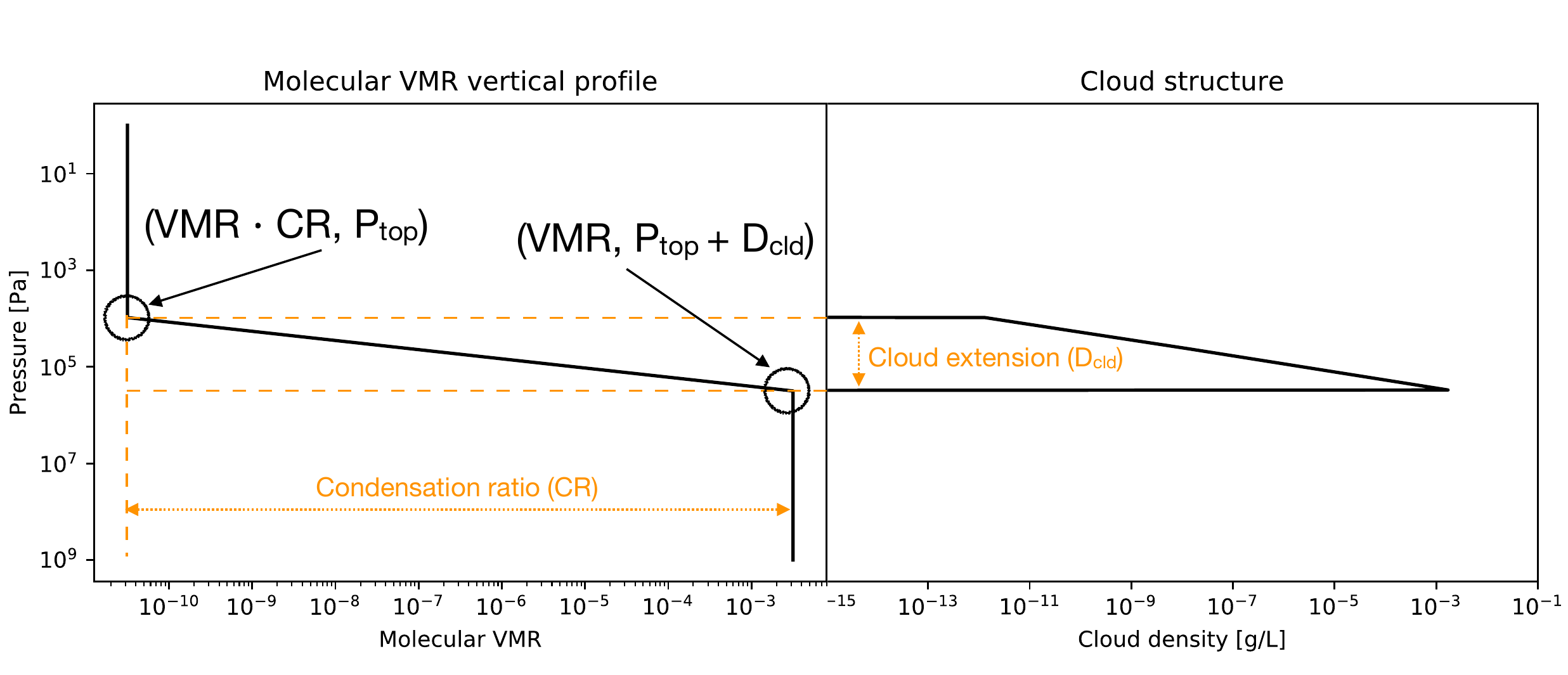}
		\caption{\textbf{Left panel:} the molecular vertical profile. In this work it could be referred to water or ammonia. Where the drop of molecular VMR occur, the cloud is present to compensate. \textbf{Right panel:} vertical profile of the cloud density relative to the VMR profile in the left panel. \label{fig:profile}}
	\end{figure*}

	In \exorelr, we wanted to preserve this causal relationship defined in \cite{Hu2019B2019ApJ...887..166H}, but, we also wanted the flexibility to change parameters to obtain a different atmospheric structure. Therefore, we ``reverse-engineered" the process by directly define a non-uniform VMR profile for water and ammonia to be used as a trigger for the calculation of the respective clouds properties. We do not consider the saturation point, we rather use four free parameters that uniquely define each VMR profile (Fig. \ref{fig:profile} left panel).
	
	\edit1{The forward model can synthesize either the albedo at a specific phase angle or the planet/star contrast ratio. In this work we focused on the albedo at a specific phase angle (afterwards referred as ``albedo" simply) as proof of concept. In this work, we fixed the phase angle to $\alpha=$60$^\circ$ for the synthesized examples (Sec. \ref{sec:impact} and \ref{sec:exop}), and to the one reported in \cite{Karkoschka1994} for the Jupiter example (Sec. \ref{sec:sol_sys}). For the retrieval process, choosing to synthesize the albedo made the gravity of the planet less significant as free parameter, which instead is important if the planet/star contrast ratio is the quantity to be retrieved (see Sec. \ref{sec:ch4_g})}

	\subsection{Free parameters space} \label{sec:par_space}
	
	We chose to design the free parameters space to include general/observable features. For this reason, we did not include the single scattering albedo ($\Bar{\omega}$), the asymmetry factor ($\Bar{g}$) and the optical depth ($\tau$) within our free parameters \citep{Lupu2016, Batalha2019}. These parameters are calculated self-consistently from other parameters since the clouds are linked to a phyical model.
	The parameters space counts at maximum 10 parameters when both water and ammonia condensates are considered. Four parameters are used to determine the VMR vertical profile of the water and four more to describe the ammonia one. As mentioned in Sec. \ref{sec:structure}, we defined the following free parameters for each of the molecular VMR vertical profile
	
	\begin{itemize}
		\item{the VMR of the molecule below the respective cloud layers;}
		\item{the P$_{top}$ as the altitude in term of pressure where the top layer of the cloud is present;}
		\item{the vertical extension of the cloud (D$_{cld}$), which quantify (in terms of difference) how much the cloud extend downward from the P$_{top}$;}
		\item{the condensation ratio (CR) that accounts for the ratio between the leftover water in the gaseous form above the cloud and the molecular VMR below it.}
	\end{itemize}
	
	
	The vertical profile of water and ammonia are defined on a pressure grid spanning from 10$^1$ to 10$^9$ Pa, from top to bottom (see Fig. \ref{fig:profile} left panel). Moving upwards, the VMR could drop due to the condensation of the relative molecular species. The drop, is modeled as a linear decrease in logarithmic space. The number of layers where the VMR drops is regulated by the P$_{top}$ and D$_{cld}$. The VMR drop is then defined:
	
	\begin{equation} \label{eq:vmr_drop}
		Log(\Delta X) = \frac{Log(X_{bot}) - Log(X_{bot} \times CR)}{N_{layers}}
	\end{equation}
	
	\noindent where $N_{layers}$ is the number of layers between P$_{top}$ and P$_{bot}$ and $X$ is the VMR. The four free parameters, previously mentioned, define uniquely the molecular vertical profile. This assumption does not create appreciable differences with the proper and consistent calculation of the cloud density profile presented in \cite{Hu2019B2019ApJ...887..166H}. In Fig. \ref{fig:profile} right panel is shown the cloud relative to the defined molecular VMR vertical profile shown on the left panel. According to \cite{Hu2019B2019ApJ...887..166H}, the cloud density has been calculated as follow
	
	\begin{equation} \label{eq:cld_den}
		\rho_{cld} = \frac{\Delta X_i \mu P_i}{RT_i}
	\end{equation}
	
	\noindent where $\Delta X_i$ is the VMR difference between two consecutive layers, $\mu$ is the molecular mean weight of the atmosphere, $P_i$ and $T_i$ are respectively the pressure and the temperature of the relative layer, and $R$ is the gas constant.
	
	Finally, the remaining two parameters are the VMR of the methane (considered constant) and the gravity acceleration of the planet.
	
	\edit1{A few challenges of the model used in this work arise from the assumptions of the model itself. The configuration used (see Sec. \ref{sec:model}) implies that the clouds are water or ammonia purely; however, this is not always the case. In the case of Jupiter, for example, the ammonia clouds are not made of ammonia solely, photochemical hazes are present, and their influence can also be appreciated in the bluest part of the reflection spectrum \citep{Weidenschilling1973, Sato1979, Karkoschka1994, Karkoschka1998}. Since the model does not include the effect of hazes yet, we did not include the data points below $0.6\ \mu$m of the Jupiter albedo in our retrieval exercise (see Sec. \ref{sec:sol_sys}). We can also expect that other cold gaseous exoplanets may have photochemical hazes in their atmosphere. It is crucial, then, to include the effects of hazes to obtain the best realistic scenario.}
	
	\edit1{In light of the mechanism of this model, other condensable species (e.g., NH$_4$SH and CH$_4$) have to be included, so that other cold gaseous planet scenarios can also be addressed (e.g., Neptune-Like planets).}
	
	\subsection{\multinest settings} \label{sec:ret_settings}
	
	The \multinest algorithm \citep{Skilling2004, Skilling2006, Sivia2006, Feroz2007, Feroz2009, Feroz2013, Buchner2014} is an established and robust method in the analysis of the free parameters space to recognize correlations and best parameters values for a model. Among its qualities, \multinest is designed to better handle multimodal posteriors. Unlike Monte Carlo Markov Chain (MCMC) algorithms, \multinest can better avoid getting stuck into a local minimum. Moreover, the calculation of the Bayesian evidence is already included in the \multinest algorithm. The evidence allows the generalization of the Occam’s razor: a theory with compact parameter space (i.e simpler) will have larger evidence than a more complicated one unless the latter is significantly better at explaining the data. We implemented this concept in our model by calculating the Bayesian factor, $\mathcal{B}$ to determine which between two models ($\mathcal{M}_{1}$ and $\mathcal{M}_{2}$) better represents the data. The Bayesian factor has been calculated as follows \citep{Trotta2008}
	
	\begin{equation}
		\mathcal{B}_{\frac{\mathcal{M}_{1}}{\mathcal{M}_{2}}} = \frac{\mathcal{P}(\mathcal{M}_{1}|\mathcal{D})}{\mathcal{P}(\mathcal{M}_{2}|\mathcal{D})} = \frac{\mathcal{Z}_{1}}{\mathcal{Z}_{2}}\frac{\mathcal{P}(\mathcal{M}_{1})}{\mathcal{P}(\mathcal{M}_{2})}
	\end{equation}
	
	\noindent where $\mathcal{Z}$ is the total Bayesian evidence of the model and $\mathcal{D}$ represents the data. Generally, we would assume that $\mathcal{M}_{1}$ and $\mathcal{M}_{2}$ are equally likely. For this reason the ratio $\frac{\mathcal{P}(\mathcal{M}_{1})}{\mathcal{P}(\mathcal{M}_{2})}$ is irrelevant for the determination of which model is better for the data. The choice is only related to the total evidence of the Bayesian sampling of the two models.
	
	For the \multinest algorithm, we choose a Gaussian as likelihood function (i.e., the standard choice). The priors for all the possible scenarios implemented in the algorithm are listed in Tab. \ref{tab:priors}. The choice of priors is fundamental for our scope as they reflect our initial knowledge of the problem. For this reason, the priors have been defined uniform among the ranges in Tab. \ref{tab:priors} to give the same probability to all the possible values. Moreover, the ranges have been defined large enough to not influence the final result of the Bayesian sampling \citep{Skilling2004, Skilling2006, Sivia2006}.
	
	Note that in the case of the 2-clouds model, the P$_{top}$ of the deeper cloud (in \exorelr the water cloud is always below the ammonia one) is instead defined to be relative to the bottom of the upper cloud and not to the top of the atmosphere. In this way, the two clouds are always separated and well distinguished.
	
	\begin{deluxetable}{c|ccc}
		\tablecaption{Priors for each scenario for the \multinest algorithm. \label{tab:priors}}
		\tablehead{
			\colhead{Parameter} & & \colhead{Cloud Models}}
		\startdata
		& Water & Ammonia & 2-cloud \\
		\hline
		$Log(VMR_{H_2O})$ & [$-12.,0. $] & [$-12.,0. $] & [$-12.,0. $] \\
		$Log(VMR_{NH_3})$ & [$-12.,0. $] & [$-12.,0. $] & [$-12.,0. $] \\
		$Log(VMR_{CH_4})$ & [$-12.,0. $] & [$-12.,0. $] & [$-12.,0. $] \\
		$Log(P_{top, H_2O})$ & [$0.,9.$] & - & [$0.,8.$] \\
		$Log(D_{cld, H_2O})$ & [$0.,9.$] & - & [$0.,8.5$] \\
		$Log(CR_{H_2O})$ & [$-12.,0.$] & - & [$-12.,0. $] \\
		$Log(P_{top, NH_3})$ & - & [$0.,9.$] & [$0.,8.$] \\
		$Log(D_{cld, NH_3})$ & - & [$0.,9.$] & [$0.,8.5$] \\
		$Log(CR_{NH_3})$ & - & [$-12.,0. $] & [$-12.,0. $] \\
		$g$ & [$10.,100.$] & [$10.,100.$] & [$10.,100.$] \\
		\enddata
	\end{deluxetable}
	
	\section{Impact of parameters to the planetary albedo spectrum}\label{sec:impact}
	
	\subsection{VMR$_{CH_4}$ and g} \label{sec:ch4_g}
	
	\begin{figure*}[]
		\plottwo{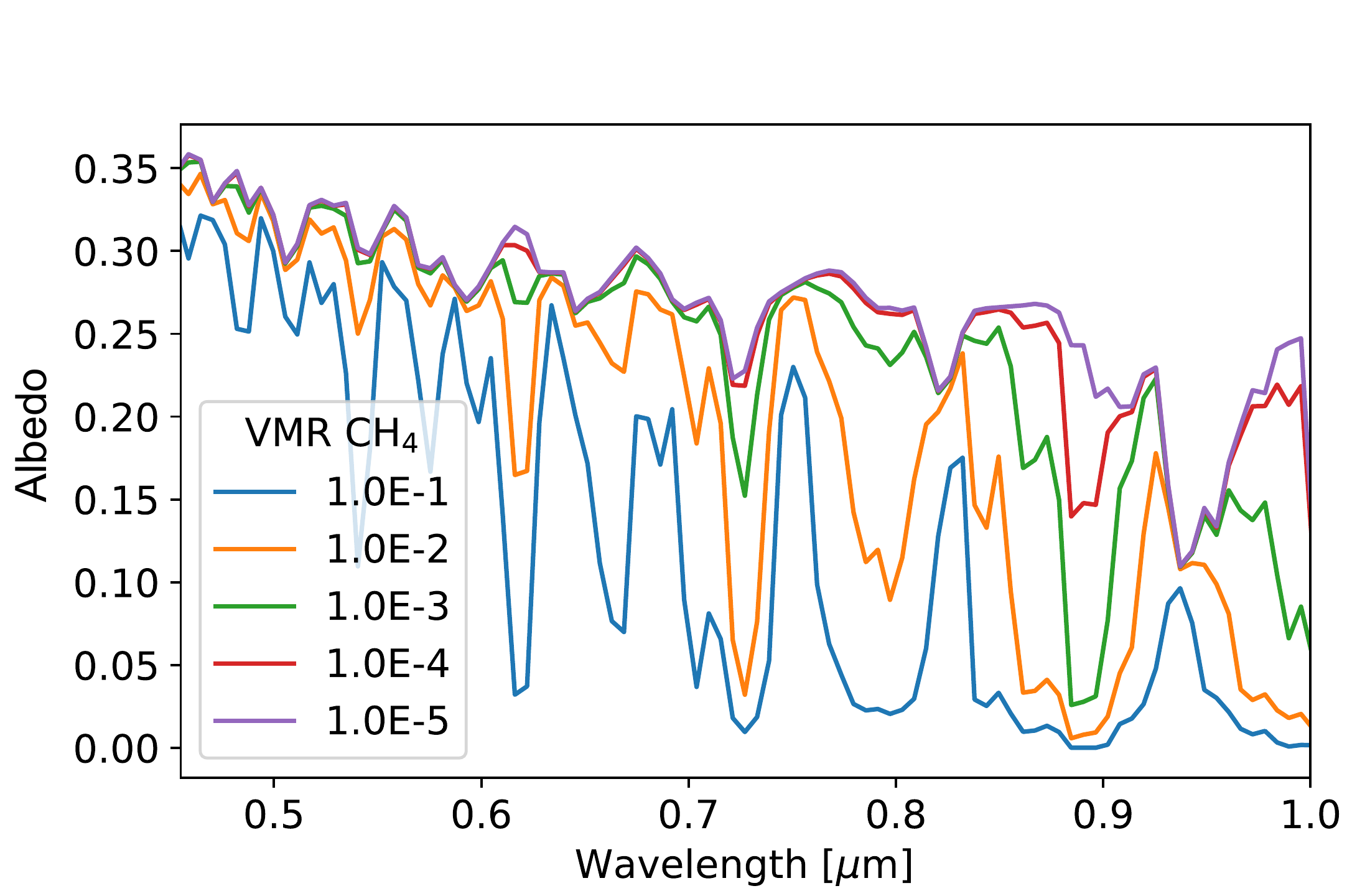}{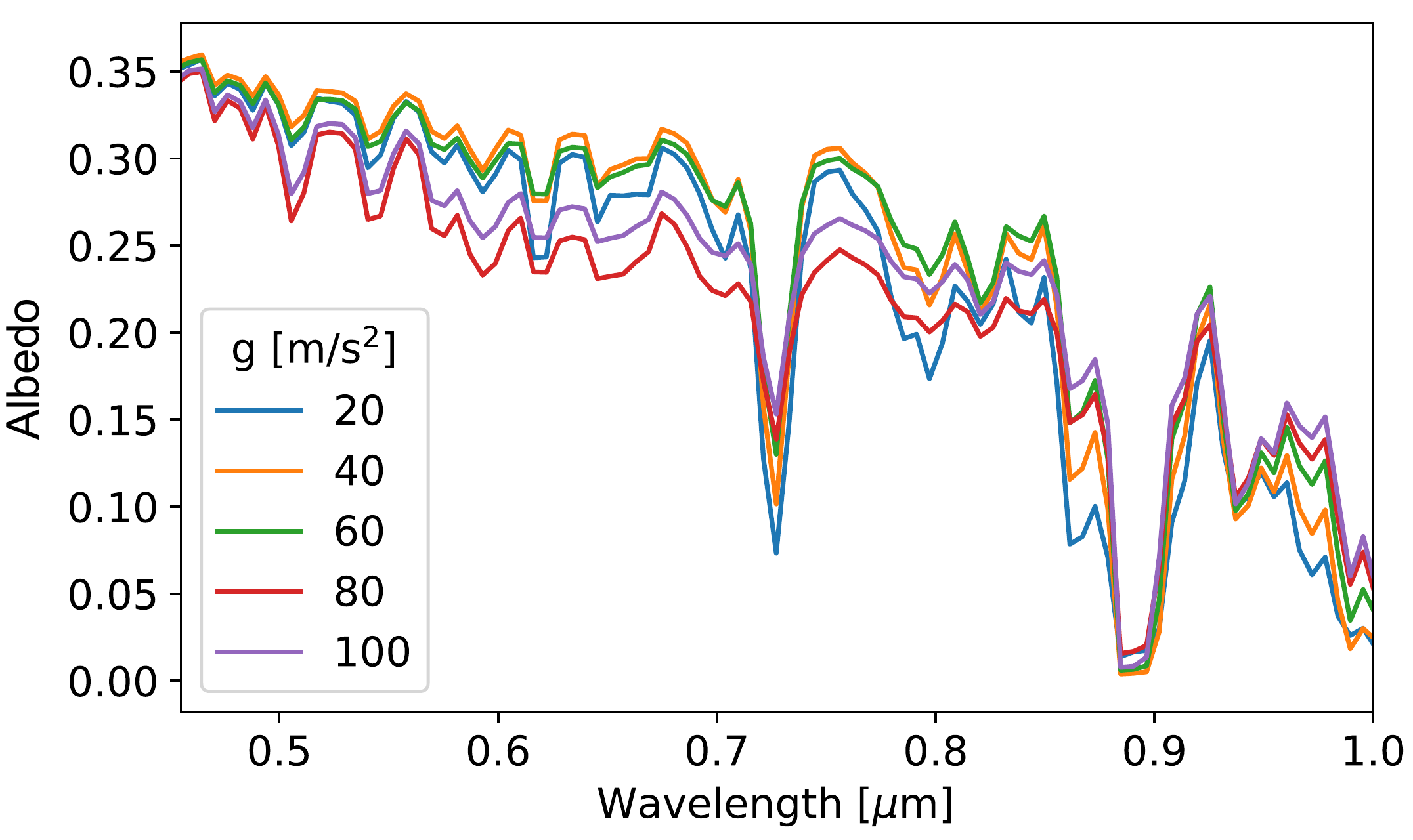}
		\caption{\textbf{Left panel:} the effect on the albedo due to the variation of the concentration of molecular methane in the atmosphere. \textbf{Right panel:} the variation of the planetary reflectivity produced by different gravity value. For these graphs the following parameters have been adopted: log(VMR$_{H_2O}$)$=$-2.5, log(VMR$_{NH_3}$)$=$-3.4, log(P$_{top, H_2O}$ [Pa])$=$4.0, log(D$_{cld, H_2O}$ [Pa])$=$5.5, and log(CR$_{H_2O}$)$=$-8.0, log(P$_{top, NH_3}$ [Pa])$=$3.0, log(D$_{cld, NH_3}$ [Pa])$=$3.60, and log(CR$_{NH_3}$)$=$-8.0, and where applicable log(VMR$_{CH_4}$)$=$-2.8, g$=$50 m/s$^2$. The spectral resolution is R=70 and the phase angle is $\alpha=$60$^\circ$. \label{fig:ch4}}
	\end{figure*}
	
	In the wavelength range within 400 and 1000 nm there are numerous methane absorption bands (see Fig. \ref{fig:ch4} left panel). The concentration of CH$_4$ affects the depth of these absorption bands. In Fig. \ref{fig:ch4} left panel, the clouds have been located at low altitude to show the methane molecular bands. The absorption can be severe with a high concentration of CH$_4$.
	
	Gravity plays a weaker role in the calculation of the albedo. Mostly, it affects the shape and depth of molecular features. For the planet studied through high-contrast imaging we may assume to know the mass of the planet, so, the gravity will give us information about the radius of the planet. Note that the effect of the gravity will be better appreciated when the planetary flux or the planet/star contrast ratio is retrieved, instead of the albedo, as it depends directly on the planetary radius.
	
	\subsection{P$_{top}$, D$_{cld}$, and CR}
	
	\begin{figure*}[]
		\plotone{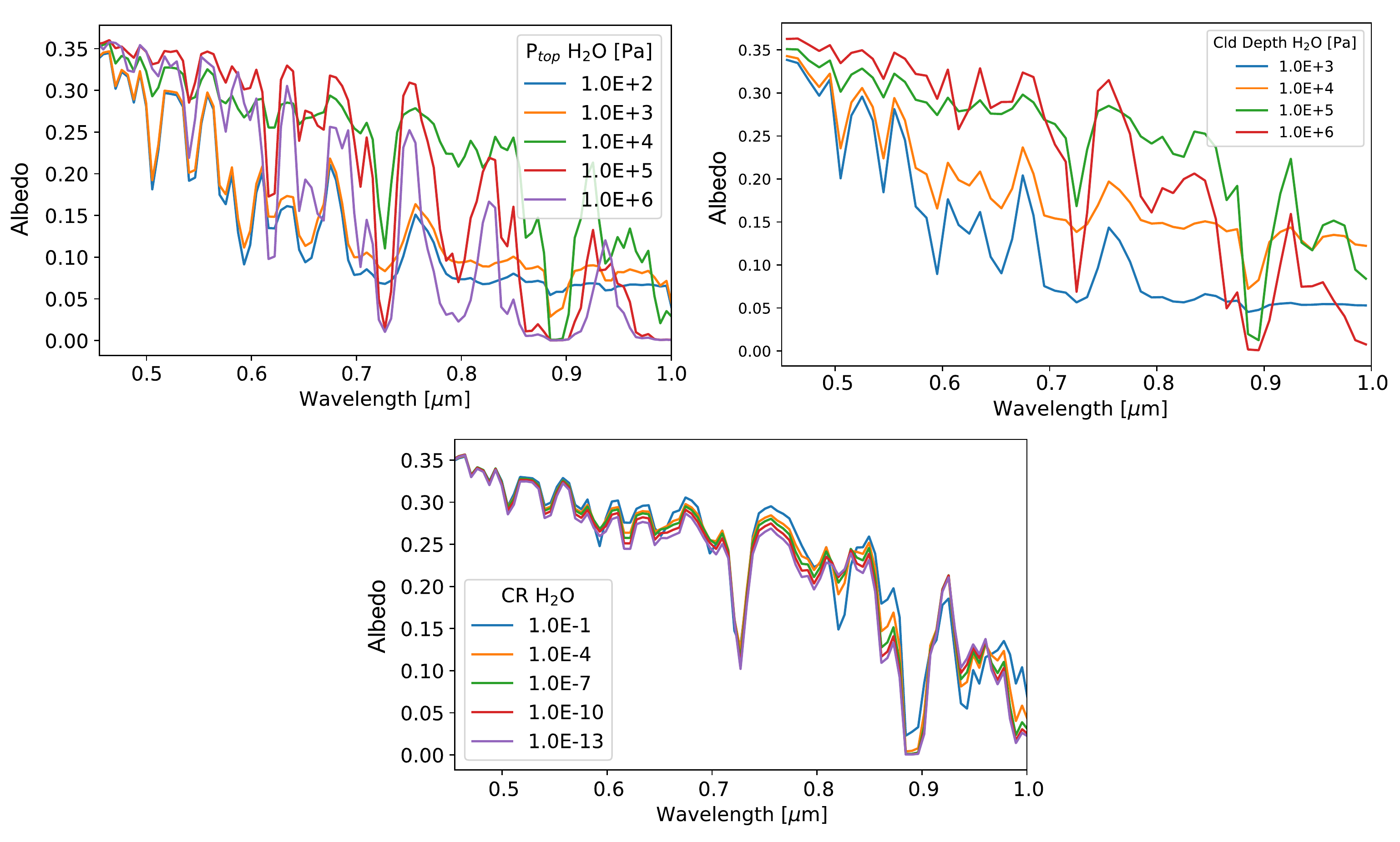}
		\caption{\textbf{Top left panel:} the effects of the variation of the albedo due to the P$_{top}$. \textbf{Top right panel:} the variation of the planetary reflectivity produced by different cloud thicknesses. \textbf{Bottom panel:} the albedo in relation to the condensation ratio. The behavior of these parameters on ammonia clouds is similar. For these graphs the following parameters have been adopted: log(VMR$_{H_2O}$)$=$-2.5, log(VMR$_{NH_3}$)$=$-3.4, log(VMR$_{CH_4}$)$=$-2.8, g$=$50 m/s$^2$, and where applicable log(P$_{top}$ [Pa])$=$4.0, log(D$_{cld}$ [Pa])$=$5.5, and log(CR)$=$-8.0. The spectral resolution is R=70 and the phase angle is $\alpha=$60$^\circ$. \label{fig:ptop}}
	\end{figure*}
	
	P$_{top}$, D$_{cld}$, and CR are the parameters that together with the molecular VMR uniquely define the cloud density vertical profile (see Sec. \ref{sec:par_space} and Fig. \ref{fig:profile}). 
	
	P$_{top}$ regulates the vertical position of the cloud and the effects on the planetary reflectivity can be seen in Fig. \ref{fig:ptop} top left panel. When the cloud is located high in the atmosphere (low pressure), the cloud density is not high enough to let the cloud be completely opaque, in this regime light pass trough and the resulted reflectivity is weak. While the P$_{top}$ increases (moving down in the atmosphere) the cloud is denser as the reflectivity increases. However, if the cloud is too deep, the molecular absorption (mostly CH$_4$) is predominant and the albedo shows deep absorption bands. 
	
	D$_{cld}$ affects the extension of the cloud in the atmosphere and it represents the vertical depth from P$_{top}$. While all other parameters are kept fixed, the cloud depth affects the cloud density and the layer where the optical depth reaches the unity (P$_{\tau = 1}$). If the cloud depth is small it means that the cloud is quite thin, letting most of the light through and resulting in a low scattering and strong molecular absorption. While the cloud extends further into the atmosphere, the cloud density increases and the cloud can scatter more light back to the space. At the same time P$_{\tau = 1}$ moves down and molecular absorption features emerge as more column abundance is present on top of the cloud.
	
	
	Since P$_{bot}$ is calculated as the sum of P$_{top}$ and D$_{cld}$ one of the two values may dominate the other, for this reason, we expect long tails in the posterior distribution that are not necessary related with the physics of the scenario, and every cases has to be taken into account carefully.
	
	The CR regulates the gradient of the cloud density from top to bottom of the cloud. The overall effect on the albedo is not significant, but it is useful to describe the vertical structure of the cloud. Moreover, it regulates the concentration of water and ammonia in the gas phase on top of the clouds, affecting in this way the absorption of such molecules.
	
	\subsection{VMR$_{H_2O}$ and VMR$_{NH_3}$}
	
	\begin{figure*}[]
		\plottwo{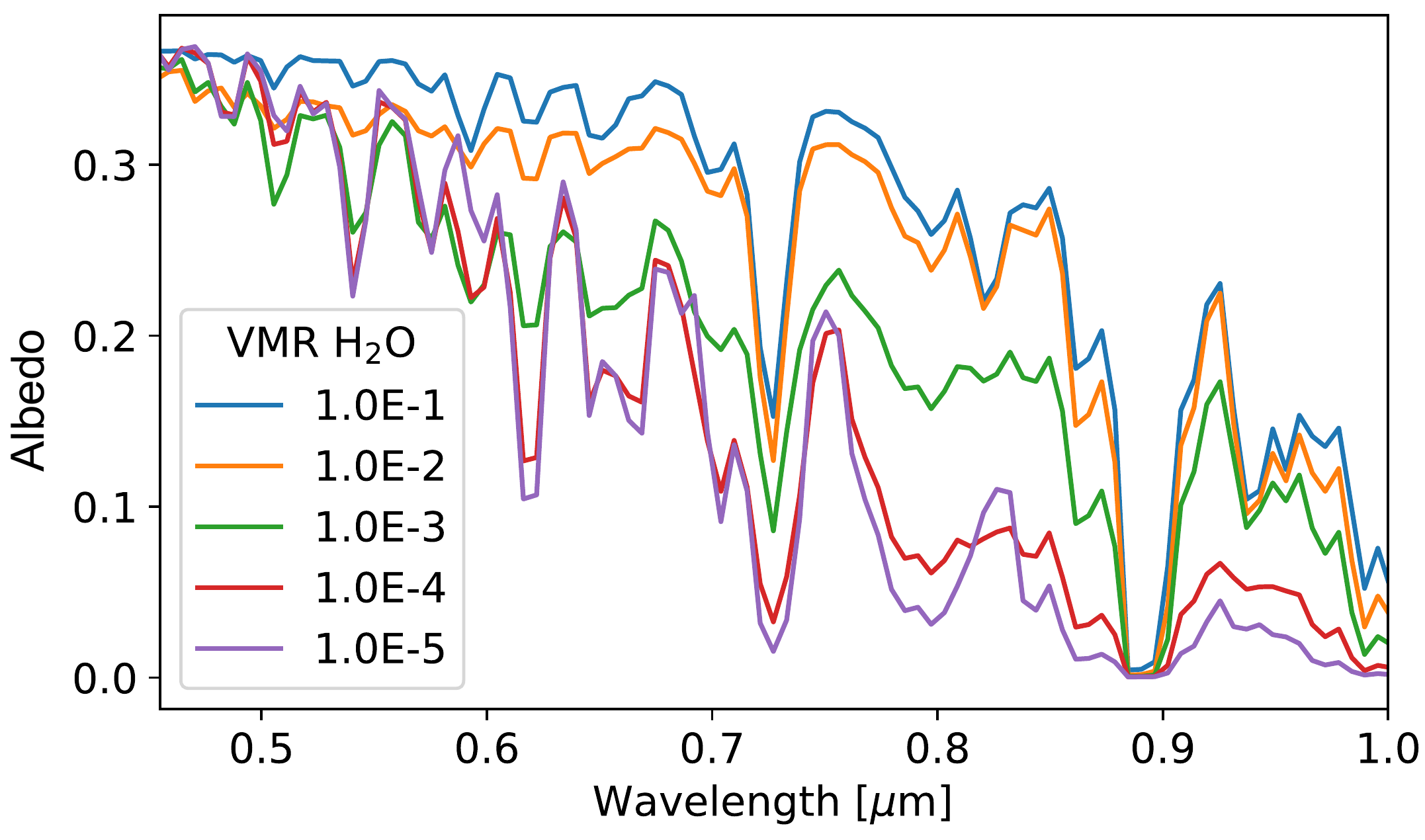}{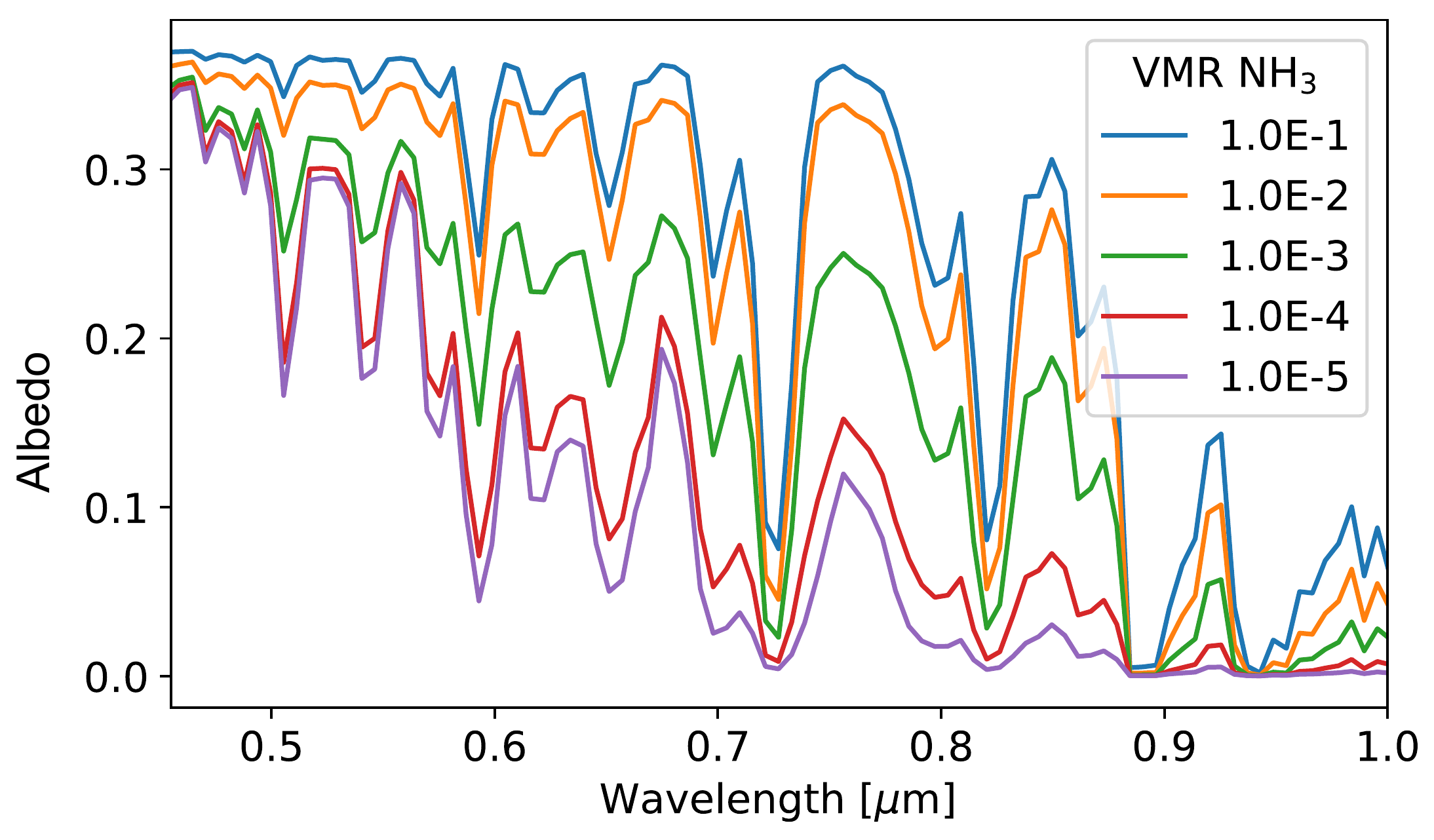}
		\caption{\textbf{Left panel:} the effect on the albedo due to the variation of the VMR of the water. Only a water cloud has been considered.  \textbf{Right panel:} Same as left panel but relative to the ammonia. Only an ammonia cloud has been included. For both graph we used log(VMR$_{CH_4}$)$=$-2.8 g$=$50 m/s$^2$. Also the P$_{top}$, D$_{cld}$, and CR are the same, but relative to the two different model: log(P$_{top}$ [Pa])$=$4.0, log(D$_{cld}$ [Pa])$=$5.5, and log(CR)$=$-8.0. The spectral resolution is R=70 and the phase angle is $\alpha=$60$^\circ$. \label{fig:vmr}}
	\end{figure*}
	
	Even though we are not able to directly measure the molecular VMR below the cloud, its effect can still be observed in the total reflectivity of the planet. The molecular VMR is directly linked with the cloud density; the higher is the concentration of that molecule, more is the material that can condense. For both ammonia and water, the behavior is indeed similar, low spectral continuum with a lower VMR and high continuum with higher VMR (see Fig. \ref{fig:vmr} and Sec. \ref{sec:below_cld}) .
	
	To calculate the albedos shown in  Fig. \ref{fig:vmr}, we tried to isolate the sole effect of the VMR to the planetary albedo. Other self-consistently calculated parameters such as the optical depth, however, may have affected the result. In the following section we show a test case where the pressure level of the optical depth equal to unity has been kept constant while changing some of the other parameters.
	
	\edit1{The impact of the H${_2}$O VMR has previously been explored on atmospheric reflected spectra \citep{MacDonald2018}. Also in their work, \cite{MacDonald2018} suggested that VMR H${_2}$O signatures impact the height of clouds and the continuum of the albedo.}
	
	\subsection{Probing deep down into the atmosphere} \label{sec:below_cld}
	
	\exorelr has been designed to reflect some key concepts explained in \cite{Weidenschilling1973, Sato1979}. Our clouds are not opaque from the top layer downward, and they are not semi-infinite clouds, they rather are finite and located in altitude. In our radiative transfer code we performs the calculations to a maximum optical depth value of $\tau_{max} = 1000$. 
	By adopting this strategy we can model photons that are absorbed or scattered by regions of the atmosphere where $\tau > 1$. This gives the possibility to model the bottom part of the atmosphere. By taking into account the relation between the vertical VMR with the cloud structure (see Eq. \ref{eq:vmr_drop} and \ref{eq:cld_den}) defined in this model, we may be able to recover the molecular VMR of some trace gasses before the depletion due to the condensation. 
	
\begin{figure*}[]
		\plotone{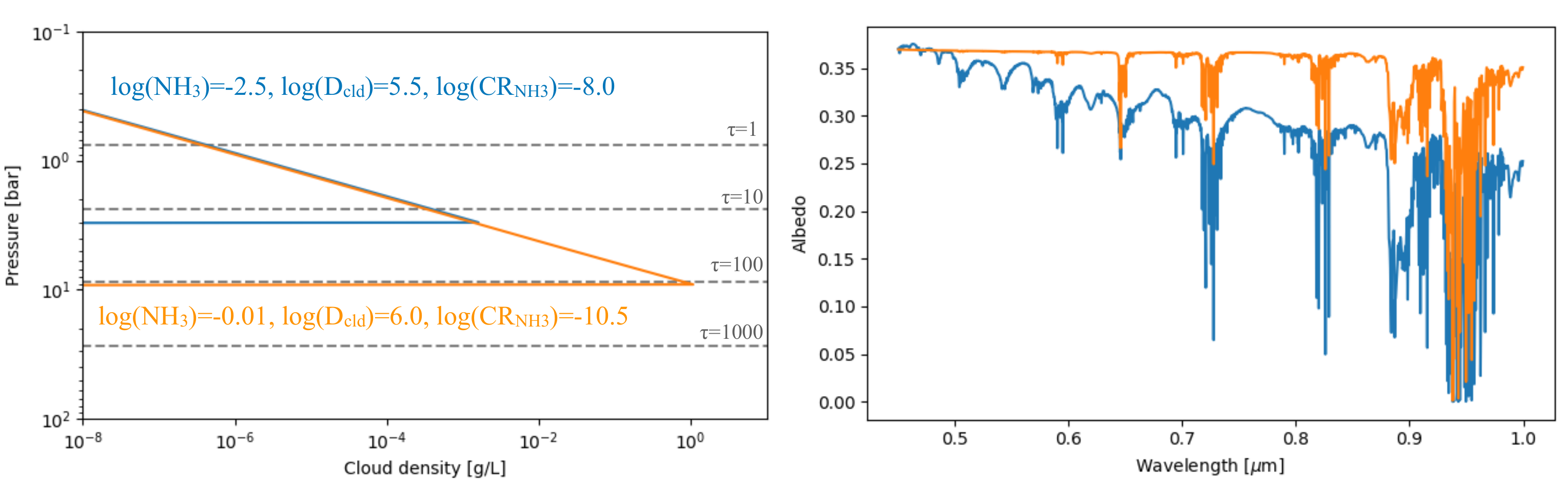}
		\caption{\textbf{Left panel:} cloud structure for the test case. The orange cloud is an extension of the blue cloud. The two models have the same values for the other parameters not reported in figure: log(VMR$_{H_2O})=$-4.0, log(VMR$_{CH_4})=$-4.0, log(P$_{top})=$-4.0, g$=$50 m/s$^{2}$. \textbf{Right panel:} the albedo resulted from considering the cloud structure on the left panel. The colors between the two panels are related. The orange cloud resulted in an higher reflectivity, however, more NH$_{3}$ is required to define such cloud. \label{fig:test_case}}
	\end{figure*}

	Fig. \ref{fig:test_case} shows a test case. We consider two clouds, one of the two is the extension of the other (the cloud described with the orange color is the extended version of the blue one). Both clouds extend below the $\tau = 1$ line (dashed line). The cloud illustrated in orange requires higher molecular concentration (in this case NH$_3$) to reach a higher value of density in the lower layer. Fig. \ref{fig:test_case} right panel shows the albedo resulted from the cloud structure scenarios. The difference between the two models is significant and it is due to higher scattering from the denser layer of the model in orange, but also higher absorption from NH$_3$ in the gas phase as it increases more than two orders of magnitude (the solely NH$_3$ absorption feature is around 0.64 $\mu$m).
	
	By using the VMR profile to define the cloud structure (Fig. \ref{fig:profile}), we can estimate the depth of the clouds as well as the VMR of the studied molecule before it condenses. For the test case and the arguments presented in this section, we expect a correlation between the D$_{cld}$ and the molecular VMR below the clouds.
	
	\subsection{Clouds: H$_2$O vs NH$_3$} \label{sec:cld_diff}
\begin{figure*}[]
		\plotone{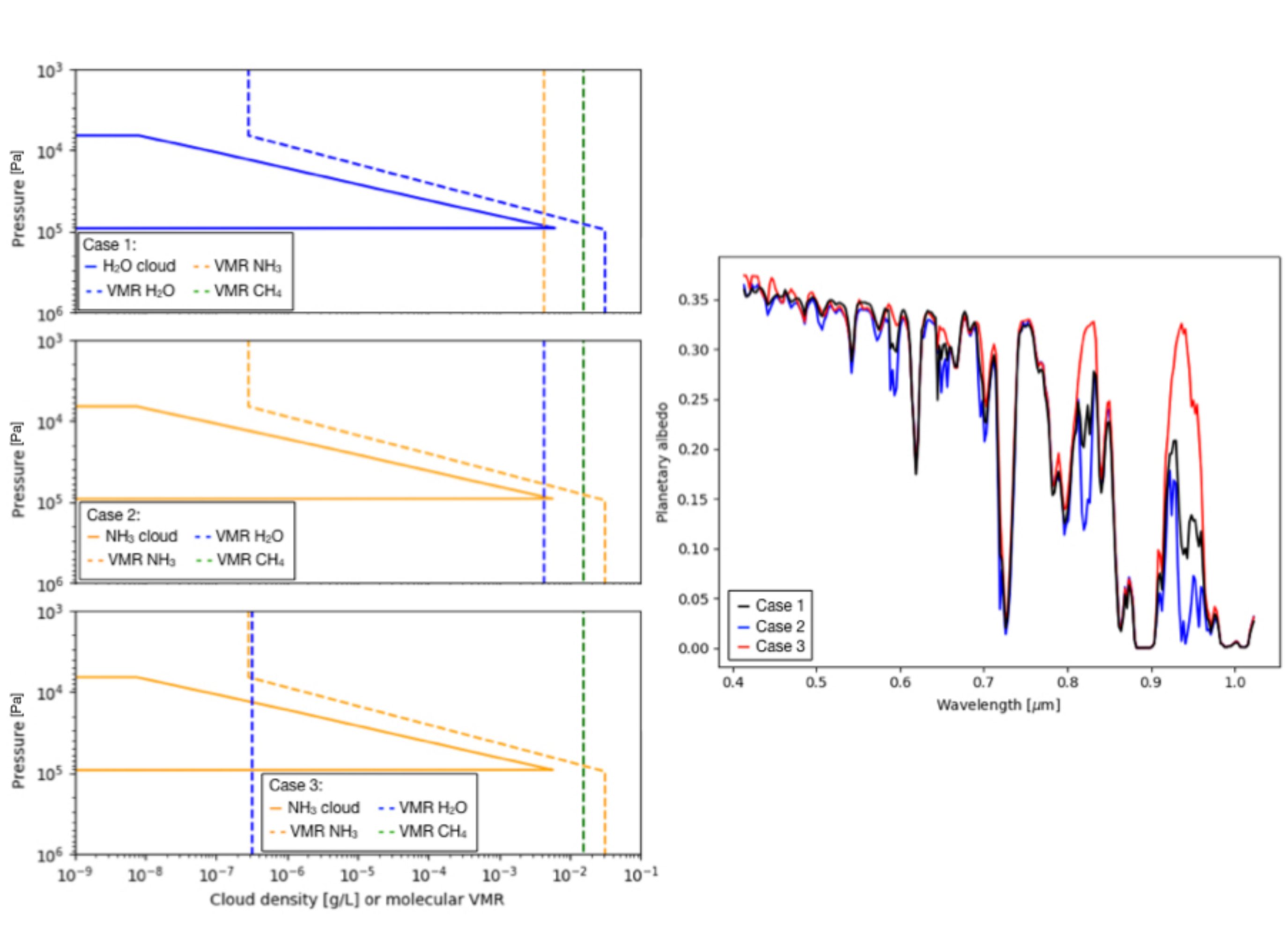}
		\caption{\textbf{Left panels:} the three different scenarios adopted to show the differences in the planetary albedo spectrum between water and ammonia clouds. The \textbf{top panel} shows a water cloud vertical profile with constant ammonia and mathane VMR. \textbf{Middle} and \textbf{bottom panels} are referred to the ammonia cloud scenarios but with different constant water VMR value. \textbf{Right panel:} the resulting atmospheric albedo spectrum for each of the three scenarios in the \textbf{left panels}. \label{fig:cloud_diff}}
	\end{figure*}

	The set-up used in this work to define the atmospheric vertical cloud distribution allows us to identify different types of clouds chemically. Depending on which VMR, either H$_2$O or NH$_3$, is modified, different types of clouds are calculated with particular cloud properties (e.g., single scattering albedo). In this section, we wanted to test if the cloud chemical composition difference can also be observed in terms of the planetary albedo spectrum. To test this hypothesis, we artificially constructed three different test cases (Fig. \ref{fig:cloud_diff}). We simulated three artificial atmospheric scenarios for a Jupiter-like gas giant.
	Firstly, we synthesized the atmospheric albedo spectrum (black curve on the right panel of Fig. \ref{fig:cloud_diff}) resulting from the presence of a water cloud at around 0.1 - 1 bar and constant VMR for NH$_3$ and CH$_4$ (top-left panel of Fig. \ref{fig:cloud_diff}). Secondly, we switched the role of ammonia and water by mirroring the values used for the first case (middle-right panel of Fig. \ref{fig:cloud_diff}). \edit1{This case is unrealistic as the condensation of NH$_3$ implies the one of H$_2$O (ammonia condenses at lower temperature), so the VMR of water should be lower in the higher part of the atmosphere. However, as proof of concept, the resulting albedo of the second scenario (blue curve on the right panel of Fig. \ref{fig:cloud_diff}) is noticeably different from the first case, mostly in the red part of the spectrum. This is because in the second case the water VMR is high across the atmosphere resulting in a stronger water absorption bands.} \edit1{Finally, we simulated a more realistic} third case (bottom-left panel of Fig. \ref{fig:cloud_diff}) in which the NH$_3$ cloud structure is the same of the one used in the second scenario, but the water is now at a realistically lower VMR (similar to the one obtained in the first case above the water cloud). Even in this scenario, the resulted albedo spectrum shows differences at longer wavelengths (red curve on the right panel of Fig. \ref{fig:cloud_diff}). The absorption due to water is now weak, and the albedo values at about 0.82 and 0.95 $\mu$m is high.
	
	By analyzing these three cases, it is then possible, in principle, to discriminate between the different scenarios and cloud structures. However, it is essential to underline that the algorithm presented in this work is used as a forward model for a Bayesian sampler that always finds the best set of parameters that produce the best model to approximate the data. For this reason, we want to point out that an intermediate scenario between cases 2 and 3 (middle and bottom right panels of Fig. \ref{fig:cloud_diff}) could be indistinguishable from case 1. However, the resulting VMR of water would be too high to be physically possible. This degeneracy can be ruled out by inferring other information to the model, e.g., the temperature of the planet, as the ammonia condenses at a lower temperature than the water.
	
	\section{Result : Exoplanet scenarios} \label{sec:exop}
	\subsection{\upsand}\label{sec:upsand}
	
	\begin{deluxetable}{c|c}
		\tablecaption{Relevant parameters used in the model for the \upsand scenario. \label{tab:ups_param}}
		\tablehead{
			\colhead{Stellar parameter} 			  & 			\colhead{$\upsilon$ And}}
		\startdata
		$R_{\star}$ (R$_{\odot}$)			&			$1.56 \pm 0.01$			\tablenotemark{1}			\\
		$T_{eff}$ (K) 								&			$6100 \pm 80$				\tablenotemark{1}		\\
		\hline
		Planetary parameters 						&	  	$\upsand$											\\
		\hline
		$M_p \times sin(i)$ (M$_{Jup}$)				&			$1.059 \pm 0.028$						 \tablenotemark{2}	\\
		$a$ (AU) 												&			$5.24558 \pm 0.00067$				\tablenotemark{2}	\\
		$e$															&			$0.00536 \pm 0.00044$	\tablenotemark{2}			\\
		$T_{internal}$ (K)									&			$110$	\tablenotemark{3}															\\
		$\alpha$ (rad)										&			$1.0472$	\tablenotemark{3}																\\
		\enddata
		\tablecomments{$^1$\cite{Butler1999}, $^2$\cite{Curiel2011}, $^3$assumed}
	\end{deluxetable}
	
\begin{figure}[]
		\plotone{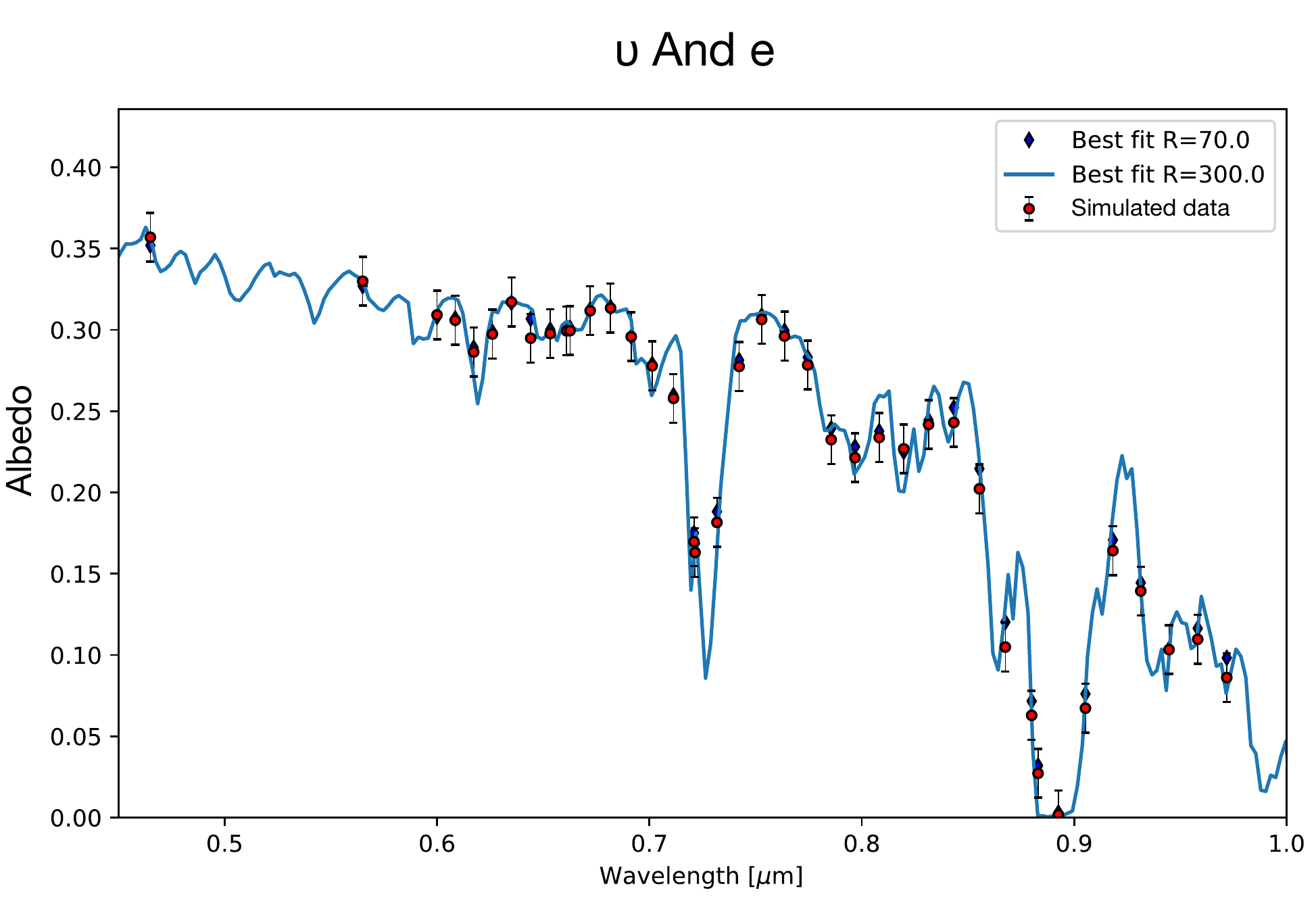}
		\caption{Albedo$-$wavelength dependence for the simulated planet \upsand. The red points represent the planetary albedo with a resolution of R$=$70 and a S/N$=$20. The \exorelr best fit model is also plotted (R$=$300 blue line, and R$=$70 blue diamond). \label{fig:ups_spec]}}
	\end{figure}

	\exorelr has been initially tested on synthesized data. We simulated \upsand scenario \citep{Butler1999, Curiel2011} which is a cold-Jupiter planet. It is one of the most Jupiter-like exoplanets found in terms of mass (M $\times$ $sin(i) =$ 1.059 M$_{Jup}$) and semi-major axis (5.2456 AU)(see Tab. \ref{tab:ups_param}). But orbiting a larger star than the Sun, it receives the same irradiation as a planet at $2.8$ AU in the Solar System. The wavelength dependence of the albedo of this planet is expected to be mostly dominated by the methane absorption and deep water cloud presence (see Fig. \ref{fig:ups_spec]}). For this reason we synthesized the data point with the values reported in the "Input" column of the Tab. \ref{tab:ups_ret}. The error bars have been calculated by considering a particular S/N (e.g., 15, 20 etc.) with respect to the difference between the albedo continuum and the bottom of the strongest methane absorption band at $\sim0.9\ \mu$m.
	
	For the information retrieval process, we set uniform priors to the parameters listed in Tab. \ref{tab:priors} relative to the water cloud case scenario.
	
	We then run \exorelr on the planetary albedo by fitting only water clouds. Fig. \ref{fig:ups_post} shows the marginalized distribution of the free parameters of the model. We have been able to not only detect and quantify the amount of methane in the atmosphere but we also recovered quantitative information about the concentration of water before its condensation level. The gravity resulted unconstrained as it does not affect the total albedo of the planet (see Sec. \ref{sec:ch4_g}). Finally, the water cloud is located between 10$^{3.45}$ and 10$^{5.55}$ Pa with a strong CR that led the VMR of water on top of the cloud to be about 10$^{-11}$. The results shown in Fig. \ref{fig:ups_post} and reported in Tab. \ref{tab:ups_ret} led to a pressure level where the optical depth reaches the unity ($\tau=1$) to 10$^5$ Pa (1 bar). By looking at the marginalized distribution we noted a weak correlation between the VMR of water and the cloud parameters as well as methane concentration with P$_{top}$, as expected. Overall, no strong correlations have been found across the free parameters space.
	
\begin{figure*}[]
		\plotone{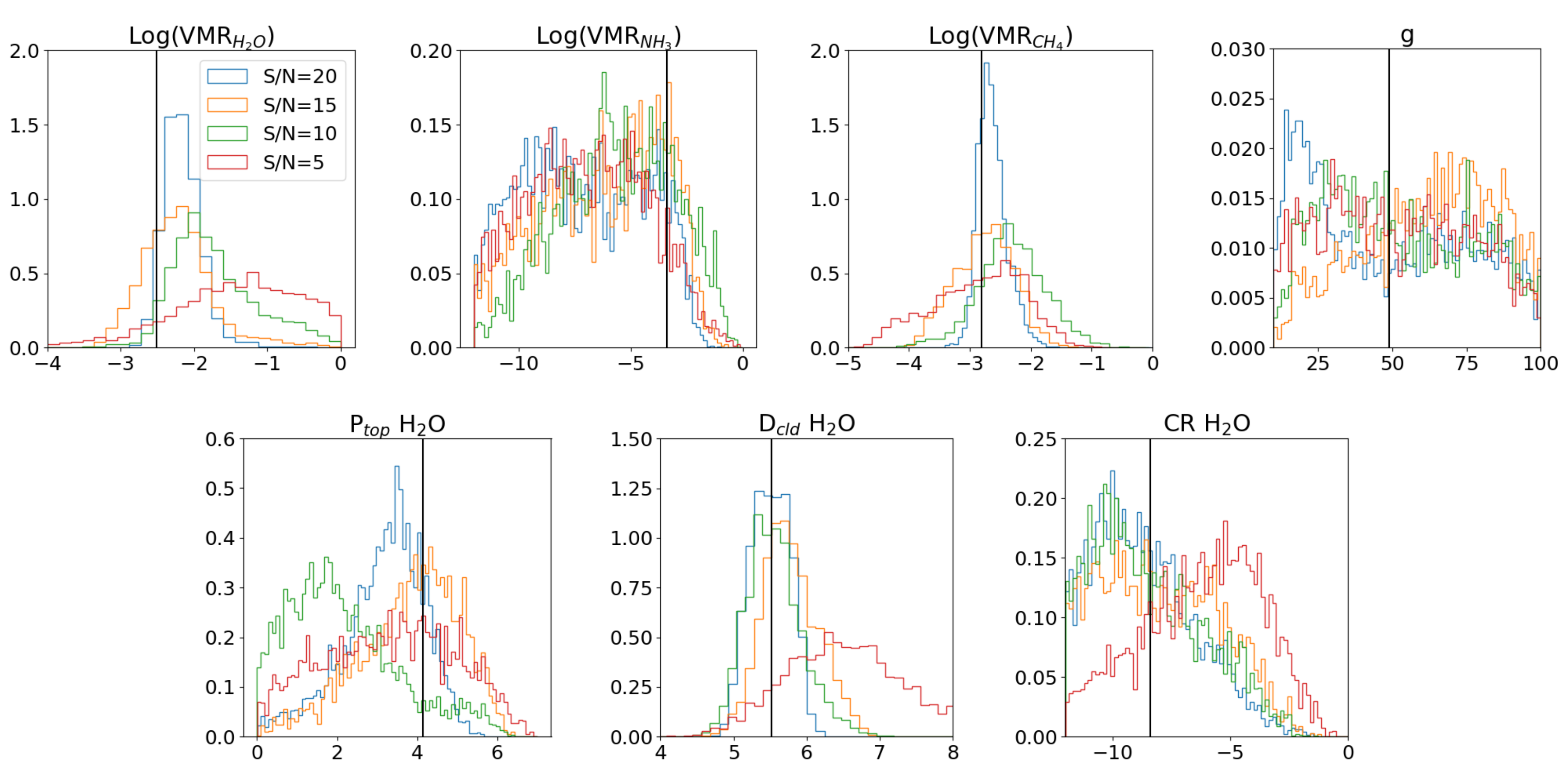}
		\caption{Marginalized distribution of the retrieved parameters at different S/N. The example is relative to the water cloud model, however, similar results are obtained with the ammonia cloud model. \label{fig:marg_dist}}
	\end{figure*}

	
	Additionally, we have used \exorelr to perform a S/N analysis to observe the performance on the retrieved parameters (see Tab. \ref{tab:ups_ret} and Fig. \ref{fig:marg_dist}). We calculated the error-bars at different S/N relative to the baseline. At S/N $=$ 5, we noticed that no constraints can be determined. At this S/N, we only have weak information about the presence of methane in the atmosphere. At S/N $=$ 10, we have a weak detection of water below the clouds and how much water have condensed, but it is not enough to constrain quantitatively these parameters. There is also a marginal quantification of methane content in the atmosphere and the cloud depth is constrained. S/N $=$ 15, presents a similar scenario with water (VMR and CR) and methane weakly constrained and the cloud depth quantified. At S/N $=$ 20, results get much better with detection of most of the parameters (except for the gravity that does not play a significant role in the albedo modulation) with most of them also constrained (e.g. H$_2$O, CH$_4$, P$_{top}$ and D$_{cld}$). NH$_3$ has narrow absorption bands in the probed wavelength range, and for this reason, it is difficult to  constrain it completely.
	
	We have also tried to retrieve information from the spectrum by excluding the bluest points of the spectrum (data points with $\lambda < 0.6\ \mu m$). We did not notice significant shifts from the results presented before.
	
	\begin{deluxetable*}{cccccc}
		\tablecaption{Retrieval results for \upsand in function of S/N. The table also report the median and 1$\sigma$ uncertainty for the marginalized distribution of the listed parameters. \label{tab:ups_ret}}
		\tablehead{
			\colhead{Parameter} & \colhead{Input} & \colhead{S/N = 5} & \colhead{S/N = 10} & \colhead{S/N = 15} & \colhead{S/N = 20}}
		\startdata
		$Log(VMR_{H_2O})$ & $-2.51$ & $-1.39^{+1.28}_{-2.58}$ & $-1.82^{+1.46}_{-0.77}$ & $-2.26^{+1.36}_{-0.80}$ & $-2.18^{+0.61}_{-0.40}$ \\
		$Log(VMR_{NH_3})$ & $-3.37$ & $-6.98^{+4.86}_{-4.55}$ & $-5.64^{+4.39}_{-5.04}$ & $-6.02^{+3.85}_{-5.48}$ & $-7.20^{+4.59}_{-4.35}$ \\
		$Log(VMR_{CH_4})$ & $-2.81$ & $-2.79^{+1.20}_{-1.59}$ & $-2.32^{+1.14}_{-1.20}$ & $-2.75^{+0.91}_{-0.96}$ & $-2.66^{+0.59}_{-0.37}$ \\
		$g$ 						 & $48.97$ & $50.29^{+44.88}_{-37.18}$ & $48.86^{+47.08}_{-33.61}$ & $64.26^{+32.65}_{-46.58}$ & $45.75^{+49.84}_{-33.55}$ \\
		$Log(P_{top, H_2O})$ & $4.14$ & $3.34^{+2.80}_{-2.98}$ & $1.91^{+3.55}_{-1.75}$ & $3.97^{+1.82}_{-2.95}$ & $3.31^{+1.53}_{-2.59}$ \\
		$Log(D_{cld, H_2O})$ & $5.52$ & $6.51^{+1.82}_{-1.50}$ & $5.53^{+0.91}_{-0.65}$ & $5.74^{+0.83}_{-0.68}$ & $5.50^{+0.48}_{-0.50}$ \\
		$Log(CR_{H_2O})$ & $-8.39$ & $-5.95^{+3.99}_{-5.36}$ & $-8.85^{+5.25}_{-2.95}$ & $-8.15^{+4.88}_{-3.60}$ & $-8.91^{+4.91}_{-2.88}$ \\
		\hline
		$ln\ \mathcal{Z}$					& 				& $68.0\pm0.3$ 					& $92.3\pm0.1$					& $102.8\pm0.2$ & $116.9\pm0.2$ \\
		\enddata
	\end{deluxetable*}
	
	\subsection{47 Uma b}\label{sec:47umab}
	
	\begin{deluxetable}{c|c}
		\tablecaption{Relevant parameters used in the model for 47 Uma b. \label{tab:uma_param}}
		\tablehead{
			\colhead{Stellar parameter} 			  & 			\colhead{47 Uma}}
		\startdata
		$R_{\star}$ (R$_{\odot}$)			&			$1.24 \pm 0.04$			\tablenotemark{1}			\\
		$T_{eff}$ (K) 								&			$5892 \pm 70$				\tablenotemark{1}		\\
		\hline
		Planetary parameters 						&	  	47 Uma b									\\
		\hline
		$M_p$ (M$_{Jup}$)				&			$2.53 \pm 0.07$						 \tablenotemark{2}	\\
		$a$ (AU) 												&			$2.1 \pm 0.02$				\tablenotemark{2}	\\
		$e$															&			$0.032 \pm 0.014$	\tablenotemark{2}			\\
		$g$ (m/s$^2$)										&			$27.8750$	\tablenotemark{3}														\\
		$T_{internal}$ (K)									&			$110$	\tablenotemark{3}																\\
		$\alpha$ (rad)										&			$1.0472$	\tablenotemark{3}																\\
		\enddata
		\tablecomments{$^1$\cite{Fuhrmann1997}, $^2$\cite{Butler1996}, $^3$assumed}
	\end{deluxetable}
	
\begin{figure}[]
		\plotone{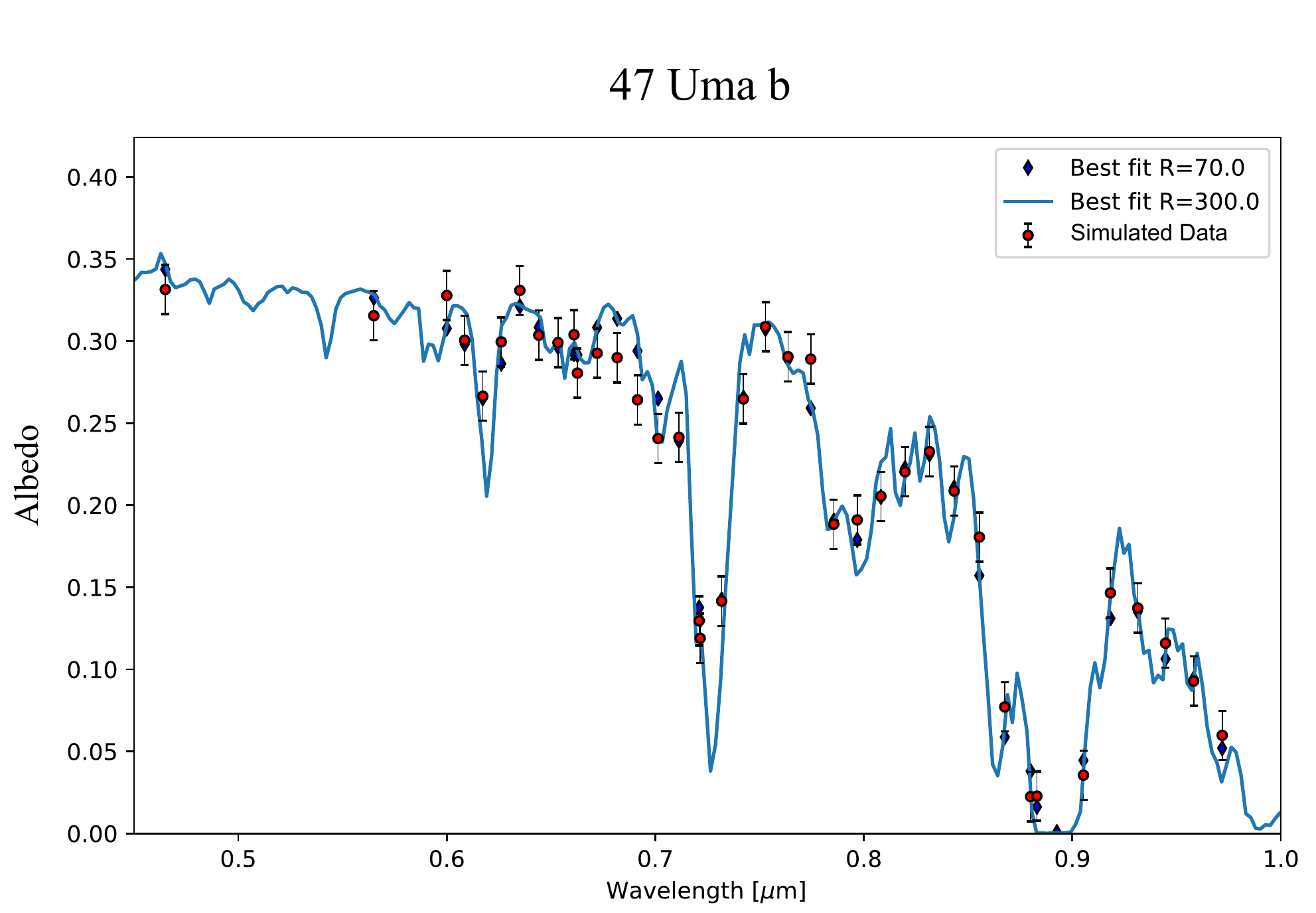}
		\caption{Best fit models to the 47 Uma b simulated data. The data have a spectral resolution of R$=$70, and the best models fit are shown at R$=$300 (solid blue line) and R$=$70 (blue diamond). \label{fig:umab}}
	\end{figure}

\begin{figure}[]
		\plotone{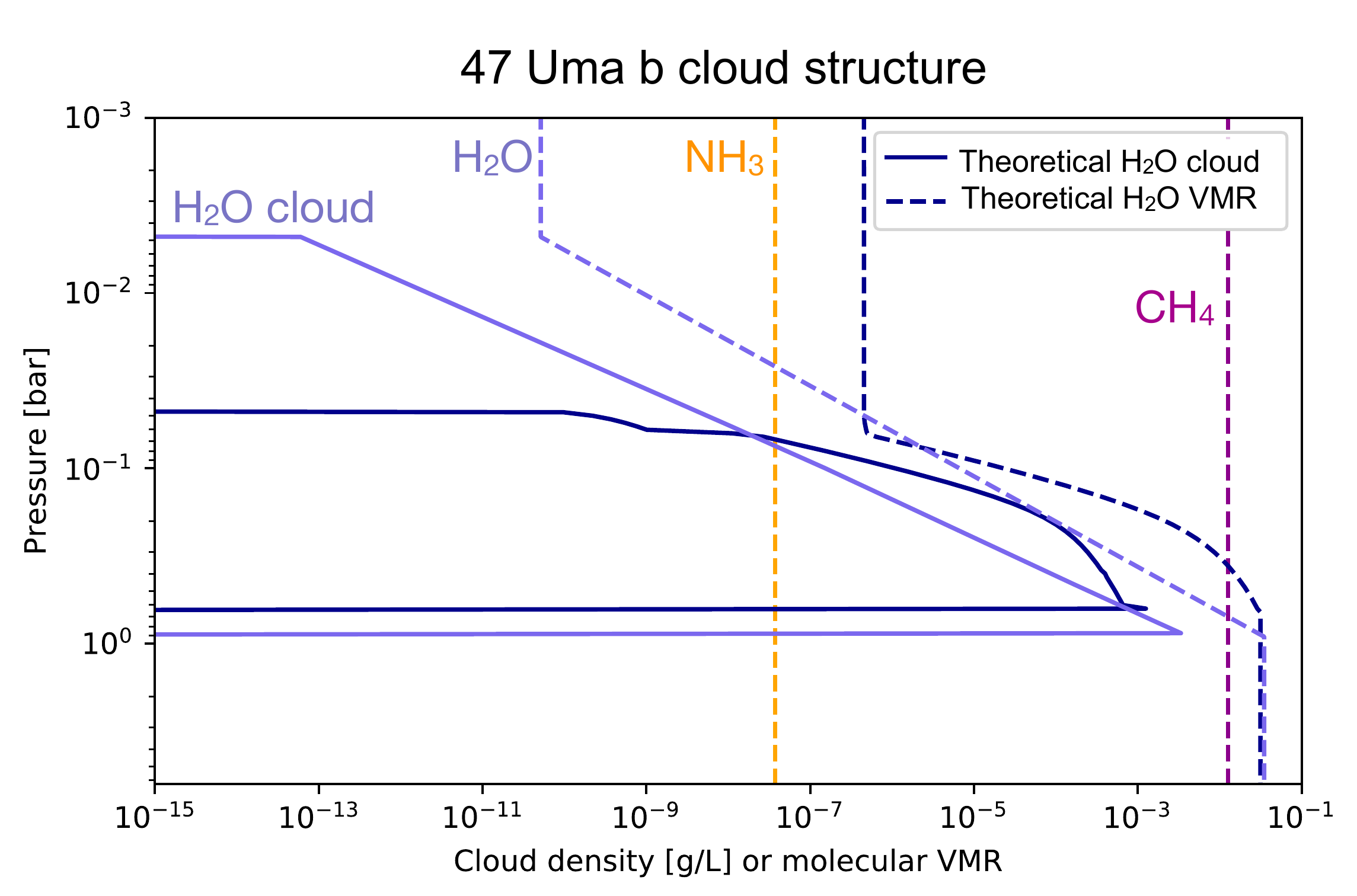}
		\caption{In dark blue, theoretical water cloud structure and theoretical VMR vertical profile of water synthesized by using our self-consistent model (\cite{Hu2019B2019ApJ...887..166H}). In light blue, the retrieved cloud structure (solid line) and the water VMR (dashed line). The VMR of ammonia (orange dashed line) and methane (purple dashed line) have been considered constant across the atmosphere.  \label{fig:uma_cld}}
	\end{figure}

	One of the exoplanets that is most likely to be observed by WFIRST for spectroscopic studies is 47 Uma b \citep{Butler1996}. It is a cold-Jupiter planet orbiting a Sun-like star (G0V) at 2.1 AU (see Tab. \ref{tab:uma_param}). With respect to the Jupiter-Sun system or the \upsand case, 47 Uma b, being closer to its host star, has got higher equilibrium temperature. In terms of cloud structure this means that the upper atmosphere of the planet is expected to contain water clouds, as opposed to the ammonia clouds typical of Jupiter \citep{Sudarsky2000}. We used our self-consistent model \citep{Hu2019B2019ApJ...887..166H} to simulate the cloud structure. The simulated scenario agrees with upper water clouds and absence of ammonia condensates.
	We interpolated the albedo in the same wavelength grid used for the \upsand scenario (Sec. \ref{sec:upsand}), we considered a S/N of 20 and we added the error-bars to the data points according to the chosen S/N. We then used our framework to fit water clouds only to the simulated data (Fig. \ref{fig:umab}). The input parameters used to synthesize the data and the result values are reported in Tab. \ref{tab:uma_ret}. 
	The marginalized distributions of the process is shown in Fig. \ref{fig:uma_post} and the theoretical and retrieved cloud structure is shown in Fig. \ref{fig:uma_cld}. The Bayesian sampling was able to retrieve and quantify the cloud extension (D$_{cld}$), the VMR of methane and the VMR of the water in the deep layers of the atmosphere. There is a weak correlation between VMR$_{CH_{4}}$ and D$_{cld}$. The median values and the errors of the marginalized distribution agree with the true value used to synthesize the albedo with the exception of the VMR of ammonia (not enough NH$_3$ bands in the wavelength range) and the condensation ratio of the water. However, the retrieved CR$_{H_{2}O}$ value ensure that on top of the clouds the water concentration drops substantially, in this way, the water absorption is absent.
	The P$_{top}$ marginalized distribution is broad as the less dense part of the cloud (i.e. the top part) is difficult to constrain, however, its retrieved distribution is fairly consistent with the input value. 
	
	\begin{deluxetable}{ccc}
		\tablecaption{Retrieval results for 47 Uma b. The table report the median and 1$\sigma$ uncertainty for the marginalized distribution of the listed parameters along side the input parameters used to synthesize the data. \label{tab:uma_ret}}
		\tablehead{
			\colhead{Parameter} & \colhead{Input} & \colhead{Results}}
		\startdata
		$Log(VMR_{H_2O})$ & $-1.50$ & $-1.46^{+1.15}_{-0.43}$  \\
		$Log(VMR_{NH_3})$ & $-2.37$ & $-7.43^{+5.14}_{-4.22}$  \\
		$Log(VMR_{CH_4})$ & $-1.80$ & $-1.90^{+0.51}_{-0.52}$  \\
		$Log(P_{top, H_2O})$ & $3.36$ & $2.70^{+1.93}_{-2.32}$  \\
		$Log(D_{cld, H_2O})$ & $4.84$ & $4.99^{+0.40}_{-0.26}$  \\
		$Log(CR_{H_2O})$ & $-4.84$ & $-8.83^{+5.63}_{-2.98}$  \\
		\enddata
	\end{deluxetable}
	
	\section{Result: Jupiter test case} \label{sec:sol_sys} \label{sec:jup}
	
	Most of the models used to explain the observation of hot exoplanets relied entirely on theoretical consideration as in the Solar System these kind of planets are not present. In the case of temperate/cold planets, we can test our model on realistic observations before applying them to exoplanetary observations. The scenario of the gaseous giant planets (Jupiter and Saturn) in our Solar System is within the scope of the model presented in this work. In this section, we present the result of the information retrieval analysis on the Jupiter recorded by \cite{Karkoschka1998}. This also gives us the possibility to present and discuss the 2-cloud vs 1-cloud model.
	
	The biggest among the Solar System's planet has been the object of deep studies to understand the composition and the structure of its atmosphere \citep{Weidenschilling1973, Sato1979, Karkoschka1994, Karkoschka1998, Wong2004, SimonMiller2001, Sato2013}. In literature, the Jupiter cloud structure is defined by NH$_3$, NH$_4$SH, and H$_2$O clouds from the highest to the lowest respectively, positioned between 1 and 10 bars \citep{Weidenschilling1973, Sato1979}. In those works, NH$_3$ and NH$_4$SH clouds are expected to be enough to describe the observations. This makes Jupiter a suitable candidate to explore the performance of our 2-cloud model (in this case we can fit only ammonia and water clouds). We adopted Jupiter's albedo measured by \cite{Karkoschka1998}. The phase angle relative to those observations is $\alpha=6.8^{\circ}$. We reduced the resolution to R$=$120 and we added error-bars in agreement with a S/N$=$20 relative to the baseline of the albedo. We then used this albedo to fed our algorithm and try to retrieve information from it. 
	
\begin{figure}[]
		\plotone{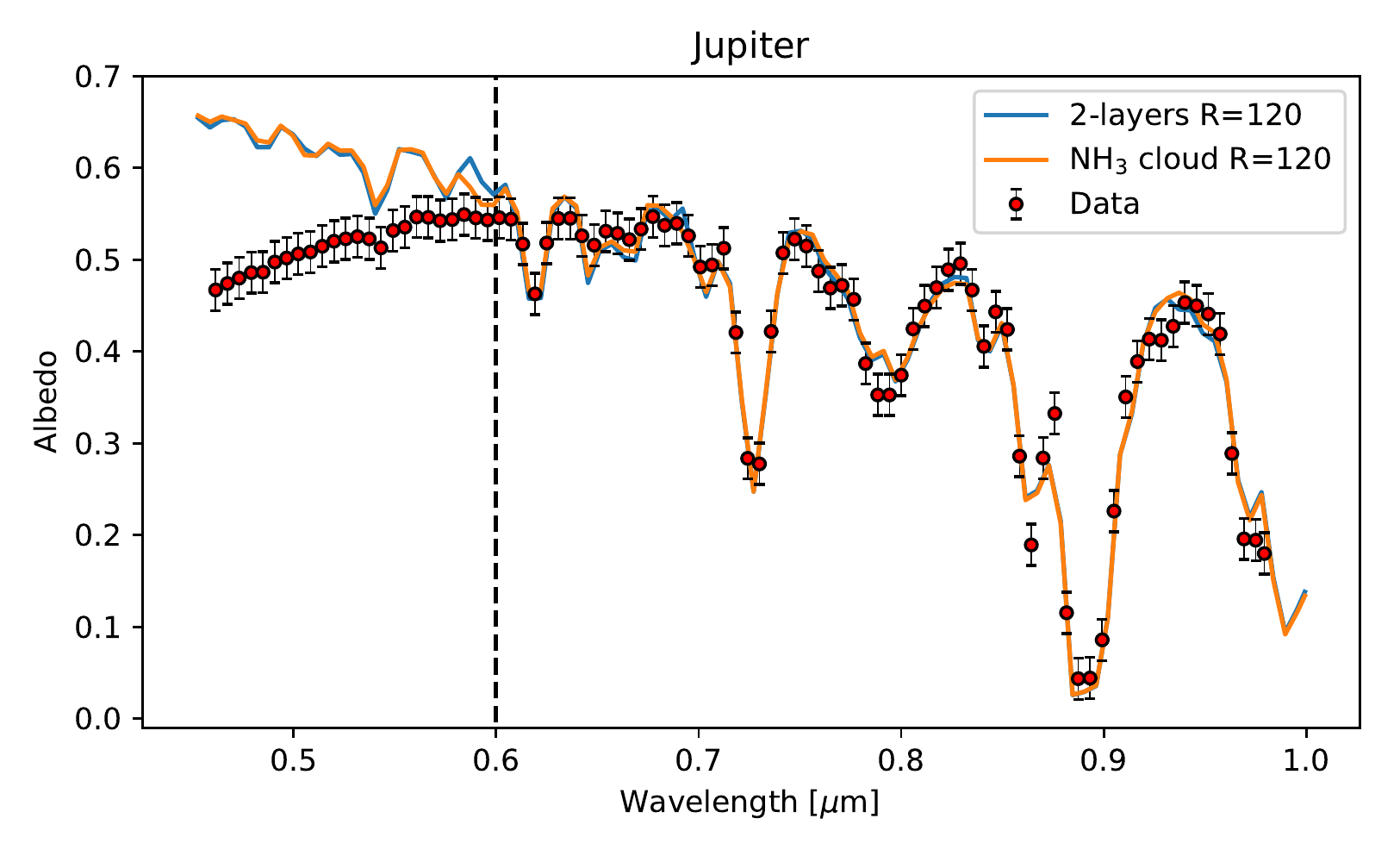}
		\caption{Best fit models to the Jupiter albedo data \citep{Karkoschka1998}. The continuous lines are relative the different cloud models: blue for the 2-clouds model and orange for the ammonia cloud model. The data have been reduced to spectral resolution R$=$120, and the best models fit are also reported at R$=$120. Data with wavelength lower than the vertical black dashed line have been excluded during the retrieval process. \label{fig:jupiter}}
	\end{figure}

	We ran \exorelr on the Jupiter's albedo with the 2-clouds model and with the ammonia cloud only model in two different instances. We calculated the Bayesian factor (see Sec. \ref{sec:ret_settings}) associated with these two models ($ln\left(\mathcal{B}_{\frac{2-clouds}{NH_3-cloud}}\right) = 1.2$), and the preference towards the 2-clouds model is not significant as the ammonia clouds alone can explain most of the spectral information (see Fig. \ref{fig:jupiter}).
	
	The results of the retrieval are reported in Tab. \ref{tab:jup_ret} and the posterior distribution is shown in Fig. \ref{fig:jup_post} and Fig. \ref{fig:jup_post2}. For completeness we show both the 2-cloud model and ammonia cloud posteriors even though the preference towards this model is not significant. The two posteriors are indeed similar in some aspects. From the retrieval, we obtained a quantification of ammonia and methane with a ratio CH$_4$/NH$_3\ \sim$ 1. The ammonia concentration, however, drops above the clouds to about $10^{-7}$. The log-concentration of methane has been recovered to $-3.65^{+0.27}_{-0.23}$ for the 2-cloud model and to $-3.54^{+0.25}_{-0.19}$ for the ammonia cloud only. The reported error-bar correspond to 1$\sigma$ confidence. If we consider 3$\sigma$, the methane concentration values are in agreement with the value reported in \cite{Wong2004}, and other retrieval work \cite{Lupu2016} in which the process has been performed on Jupiter data taken from \cite{Karkoschka1994}. We found a multi-modal solution for the concentration of water, probably due to the not high significance of the 2-cloud model. The depth of the water cloud is not significant, making the cloud too thin, which is the reason why there is no much difference between the 2-cloud model and the NH$_3$ cloud model.
	
\begin{figure}[]
		\plotone{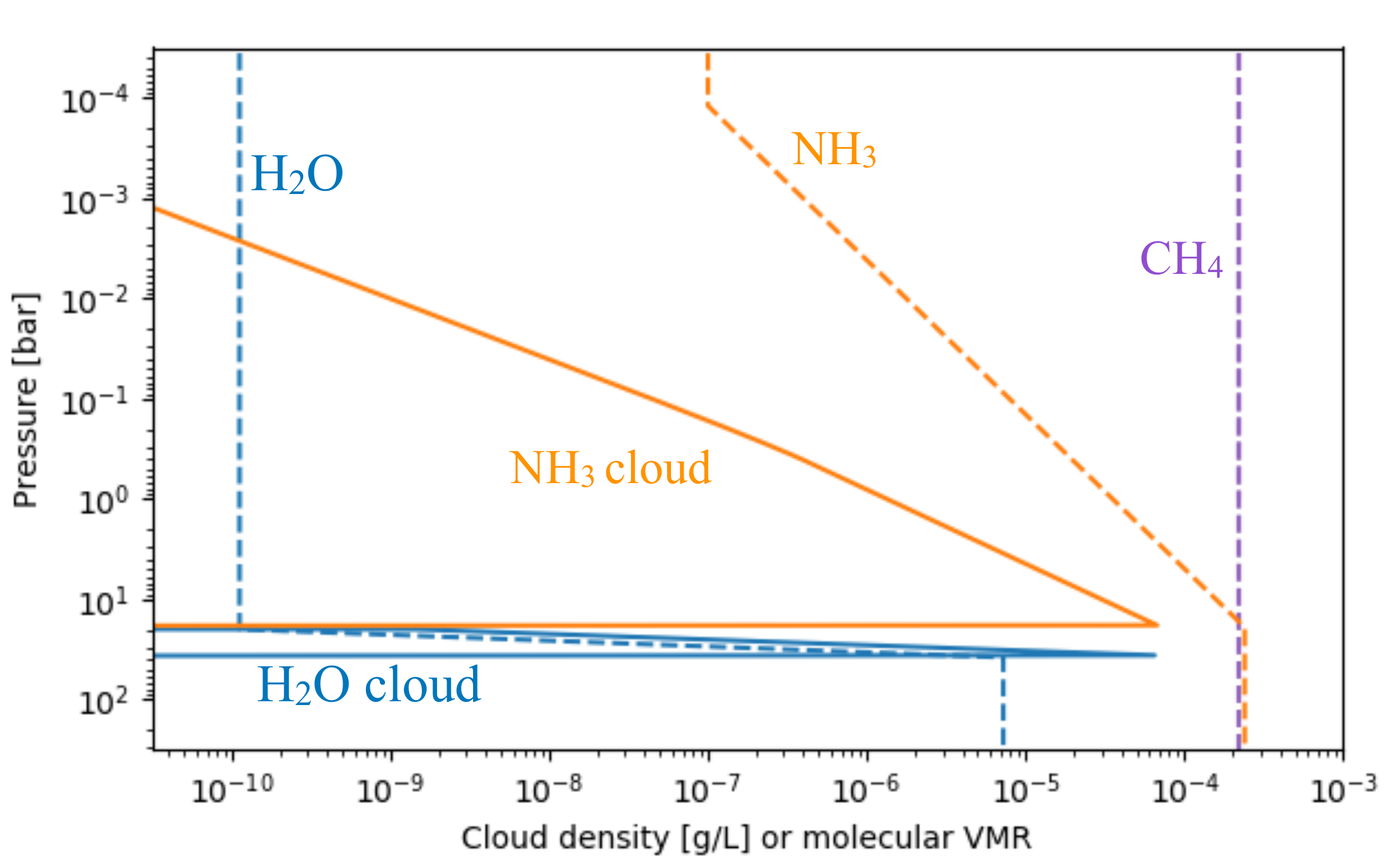}
		\caption{Retrieved atmospheric vertical profile of Jupiter. The values used to compute this graph are relative to the maximum likelihood of each parameter shown in the posterior distribution (Fig. \ref{fig:jup_post}). The volume mixing ratios of the trace gasses are represented with the dashed lines, while the clouds are represented with continuous lines. \label{fig:jup_atm}}
	\end{figure}

	The clouds position (see Tab. \ref{tab:jup_ret} and Fig. \ref{fig:jup_atm}) retrieved is in general agreement with the theorized atmospheric structure of Jupiter being between 1 and 20 bars \citep{Weidenschilling1973, Sato1979}. However, the actual structure is much more complex than the one that we obtained, with NH$_4$SH clouds and different haze layers \citep{West1986}.
	
	We noticed however, some correlations among the free parameters; the strongest is the one between the retrieved value of ammonia below the clouds and the concentration of methane. This is a consequence of the correlations of these two parameters with the depth of the ammonia cloud. Both the abundance of ammonia and the depth affect the density of the clouds (see Sec. \ref{sec:below_cld}). The correlation between the methane VMR and the cloud position is well known also in previous works \cite{Irwin2015, Lupu2016}.
	
	\begin{deluxetable}{ccc}
		\tablecaption{Retrieval results for Jupiter. The table report the median and 1$\sigma$ uncertainty for the marginalized distribution of the listed parameters. \label{tab:jup_ret}}
		\tablehead{
			\colhead{Parameter} & \colhead{2-cloud} & \colhead{$NH_3$ cloud}}
		\startdata
		$Log(VMR_{H_2O})$ & $-4.63^{+3.29}_{-2.34}$ & $-6.64^{+0.43}_{-4.22}$\\
		$Log(VMR_{NH_3})$ & $-3.65^{+0.27}_{-0.21}$ & $-3.54^{+0.24}_{-0.22}$\\
		$Log(VMR_{CH_4})$ & $-3.65^{+0.27}_{-0.23}$ & $-3.54^{+0.25}_{-0.19}$\\
		$Log(P_{top, NH_3})$ & $1.60^{+2.51}_{-1.50}$ & $1.38^{+2.35}_{-1.18}$ \\
		$Log(D_{cld, NH_3})$ & $6.32^{+0.45}_{-0.34}$ & $6.19^{+0.25}_{-0.26}$ \\
		$Log(CR_{NH_3})$ & $-3.49^{+2.01}_{-3.35}$ & $-3.64^{+1.89}_{3.40}$ \\
		$Log(P_{top, H_2O})$ &  $3.67^{+3.22}_{-3.40}$ & \\
		$Log(D_{cld, H_2O})$ &  $3.72^{+3.06}_{-3.48}$ & \\
		$Log(CR_{H_2O})$ &  $-5.37^{+4.17}_{6.15}$ & \\
		\hline
		$ln\ \mathcal{Z}$& $135.8\pm0.1$ & $134.6\pm0.1$ \\
		\enddata
	\end{deluxetable}
	
	
	\section{Discussion} \label{sec:discussion}
	
	Reflection spectroscopy is an emergent topic, and for this reason, different retrieval models have been proposed and tested. Regarding cold gaseous planets, a few models have been published \citep{Lupu2016, Nayak2017, Batalha2019, Lacy2019}. One of these model \citep{Lacy2019} is inspired by empirical observation but the free parameters are not necessary linked to physical quantities. The other models agree on having the single scattering albedo ($\Bar{\omega}$), the asymmetry factor ($\Bar{g}$) and the optical depth ($\tau$) within their free parameters as the clouds model is not linked to a physical model of cloud structure (e.g. particle size and chemical identity of cloud constituents are not taken into account).
	
	In this work we wanted to follow some of the other models' aspects and eventually propose a different point of view. For the interpretation of hot-Jupiters observations, we could not get help from observations in the solar system as this lacks this type of planets. For cold gaseous planets, it is different; we can observe them closely with much more details and take inspiration for developing general models.
	
	\cite{Weidenschilling1973} and \cite{Atreya1999} have successfully predicted the bulk cloud structure of Jupiter by considering water and ammonia as condensable species. This fact has been implemented in our model and we made a distinction between water and ammonia clouds as their physical properties are linked to the relative non-uniform molecular VMR vertical profile. However, the distinction between the two clouds given by the information retrieval of the spectrum depends on the case. The VMR of water and ammonia in the atmosphere is directly linked with the density of the corresponding cloud (Eq. \ref{eq:cld_den}); the more water or ammonia is present in the atmosphere, the more dense a cloud can be, affecting in this way the optical properties of the cloud itself.
	Our cloud model is linked with a physical model that calculates cloud density and particle size. As presented in the literature, this is likely to bring a correlation between the VMR of methane and the cloud position \citep{Irwin2015, Lupu2016}. However, there might be cases in which this correlation is not significant as highlighted by \cite{Hu2019B2019ApJ...887..166H}, it depends on a combination of S/N, spectral resolution, and particular combinations of cloud position and methane VMR.
	
	
	
	\subsection{The role of P$_{top}$ and D$_{cld}$}
	Unlike the common definition of P$_{top}$ \citep{Irwin2008, Madhusudhan2009, Benneke2012, Waldmann2015B2015ApJ...813...13W, Waldmann2015B2015ApJ...802..107W, Lupu2016, Feng2018, Batalha2019}, in the algorithm presented here, it does not play a central role. The P$_{top}$ regulates the least dense part of the cloud where not much scattering is happening. This, in part, explains why we observe broad posterior of P$_{top}$ in our marginalized histograms. Most of the scattering happens at the bottom of the cloud where it is denser. D$_{cld}$ is the parameter related to the lower part of the cloud where most of the scattering happens. Moreover, most of the time the pressure value at which the optical depth reaches the unity (P$_{\tau=1}$) is close to the bottom of the cloud. Finally, since the P$_{bot}$ is defined as the sum of P$_{top}$ and D$_{cld}$, most of the time D$_{cld}$ will dominate the summation making the posterior of P$_{top}$ broader towards lower value. To compare our work to those in the literature D$_{cld}$ is the parameter in which the attention should be focused on.
	
	\subsection{Jupiter results}
	Even though our model is inspired by solar system observations and by \cite{Weidenschilling1973, Sato1979} works, there are assumptions and simplifications that creates differences between the literature and our models. The Jupiter's cloud structures theorized in literature assume the presence of hazes and multiple (even more than two) cloud layers of different molecular species. In our model, instead, we did not include hazes. We have not modeled the condensation of NH$_4$SH, which could be necessary to have a better fit Jupiter's and other scenarios of reflected spectra. 
	
	The results we obtained for Jupiter (Tab. \ref{tab:jup_ret}, Fig. \ref{fig:jupiter}, \ref{fig:jup_atm}, and \ref{fig:jup_post}) show that a single cloud layer can be sufficient to explain the albedo modulation. However, since we have the sensibility to different cloud species, using a 2-cloud model configuration with NH$_3$ and NH$_4$SH condensates and the presence of hazes, may have a better outcome than the water cloud that does not really contribute to the albedo modulation. The overall position of the NH$_3$ cloud reflects, however, the clouds position reported in \cite{Weidenschilling1973} and also the position of clouds measured by planetary missions \citep{West1986}. We want to point out that even if the theoretical values are used in the fully consistent model \exorel \citep{Hu2019B2019ApJ...887..166H} and a theoretical clouds structure is considered, the calculated albedo, while it matches with Jupiter in its bulk part, cannot sufficiently account for the methane weak bands. This suggests that further effects need to be taken into account \citep{Hu2019B2019ApJ...887..166H}. This may also explain why the concentration of methane has been underestimated by our model. However, we would like to point out that even though these simplification have been adopted, our recovered free parameters values agree with the literature works within the 3$\sigma$ confidence.
	
	Bayesian samplers are designed to explore the parameters space to find the solution that best approximates the data. The result of this process will closely reflect the reality only if most of the effects, that take place in the process under study, are taken into account. In this sense, hazes, other absorbers, and cloud species may be required for future studies.
	
	\subsection{Water vs ammonia clouds}
	In Sec. \ref{sec:structure} we described the step required for the atmospheric structure to be constructed. We differentiated water clouds from ammonia clouds by considering different volume mixing ratio vertical profile for the two molecules and different particle size (as we used the mean molecular mass for the calculation). Also the opacities of the two molecules are different and this creates a further distinction between the two cloud species when the single scattering albedo is calculated.
	
	In Sec. \ref{sec:cld_diff} we synthesized three different scenarios to try to distinguish between water and ammonia clouds. However, whether or not we are able to distinguish between the two species with this algorithm, will depend on a case by case. By combining information from the Bayesian factor, the expected temperature of the planet and the use of the self consistent model, we might be able to discriminate between the two cloud species.
	The results of the Bayesian sampling could indeed be compared with a self consistent model and see if the two outcomes agree with each other.
	
	\subsection{Implication of constraining the molecular VMR below the clouds}
	Direct quantification of molecular abundances below the cloud deck has been a difficult task. Most of the models in the literature, that interpret atmospheric spectra, do not quantify parameters below P$_{\tau=1}$ by design. In our work, we tried to link the presence of condensates to the variation of molecular concentration. This gave us the freedom to fit the concentration value below the clouds. Essentially, we are assuming that for a certain condensate to be present (defined by density, particle size, and extension, see Sec. \ref{sec:below_cld}) a particular non-constant volume mixing ratio vertical profile is required. This assumption may have an important implication: in the cold gaseous planets scenario, it could help in improving atmospheric modeling and detect the presence of water and ammonia unseen in direct measurements. This behavior is also embedded in our algorithm as the VMR of such molecules on top of the clouds drops drastically making them almost undetectable.
	
	\edit1{\subsection{Spectral noise realization}
	In this work, we showed a novel approach on modeling and retrieving chemical abundance and cloud information from cool gaseous giant planet spectra. For this reason, we focused on the description and performances of the model without stressing the aspect of the spectral noise. In the retrieval exercises presented in this work (Sec. \ref{sec:exop} and \ref{sec:sol_sys}), we added the error-bars to the spectral data point by calculating the average albedo across the wavelength range and scaling it by the chosen S/N. This may led to underestimate the retrieval error and to introduce biases. Previous works \citep{Lupu2016, Feng2018} have shown that accounting for a random noise increases the uncertainty of the retrieved values significantly as the data points are extracted from a Gaussian distribution that changes the value of each measurement away from a simple model mean. In this context, the results presented in this work are optimistic, and including a random noise to the data points would weaken the constraint on retrieved values at high S/N (20 or 15) and completely fade out any quantitative detection at lower S/N (10 or 5).}
	
	\section{Conclusion} \label{sec:conclusion}
	In this work, we presented \exorelr, our novel Bayesian inverse retrieval algorithm for exoplanetary reflected light spectra. The gas giants' albedo (key ingredient of reflection spectroscopy) in the visible and near-infrared wavelength is mostly affected by cloud scattering and molecular absorption from H$_2$O, NH$_3$, and CH$_4$. We used a non-uniform VMR vertical profile of water and ammonia to construct water and ammonia clouds. Compared to previous retrieval models of reflected light spectra, \exorelr enforces the causal relationship between the gas abundance and the corresponding cloud density. Since \exorelr calculates the single scattering albedo, the asymmetry factor and the optical depth consistently, it employs a set of free parameters that define the non-uniform VMR of water and ammonia, e.g., the cloud depth (D$_{cld}$) and the VMR below the clouds (see Fig. \ref{fig:profile}). We presented the performances of our model with two exoplanetary test cases: \upsand and 47 Uma b which are candidates to be observed and characterized by the upcoming WFIRST mission. Finally, we have run our algorithm on a realistic case by trying to analyze the Jupiter albedo. 
	
	The key results of our work comprise: the evidence of cloud presence and position estimation, physical characterization of clouds (cloud density and particle size profiles), possibility to determine cloud chemical constituent (distinction between water and ammonia clouds), quantification of methane concentration and possible indirect quantification of the VMR of condensable molecules below the cloud.
	
	
	The retrieval exercises presented in this paper show that the reflected light spectra expected to be recorded by future space missions should be sufficient to put meaningful constraints on the presence of clouds and the abundance of methane. This conclusion is validated by the Solar System test case, where we fit a relatively simple model to the Jupiter planetary atmosphere known to be complex. If the S/N is high enough, reflected light spectroscopy may help us quantify the cloud extension, which reflects the position of the clouds, as well as the below-cloud concentration of the gas responsible for the presence of the cloud itself.
	
	For this initial instance some approximations have been made (e.g., log-linear condensation of water and ammonia, and absence of hazes).	In this work, we have focused on the possibility of retrieving parameters about the atmospheric characteristics of cold gaseous planets. In a future work this algorithm will be further developed to also include temperate/cold rocky planet scenarios with H$_2$- and non H$_2$-dominated atmospheres. 
	
	\section*{Acknowledgments}
	The authors thanks Dr. Gra\c{c}a M. Rocha and Dr. Sergi R. Hildebrandt for helpful discussions, materials, and encouragement in the preparation of this manuscript. This work was supported in part by the NASA WFIRST Preparatory Science grant \#NNN13D460T, and NASA WFIRST Science Investigation Teams grant \#NNN16D016T. This research was carried out at the Jet Propulsion Laboratory, California Institute of Technology, under a contract with the National Aeronautics and Space Administration. 
	
	{	\small
		\bibliographystyle{apj}
		\bibliography{ExoReL.bib,ExoReL2.bib}
	}	
	
	\appendix
	\section{Posterior distributions}
	
	\subsection{\upsand - water cloud}
	\begin{figure}[!h]
		\plotone{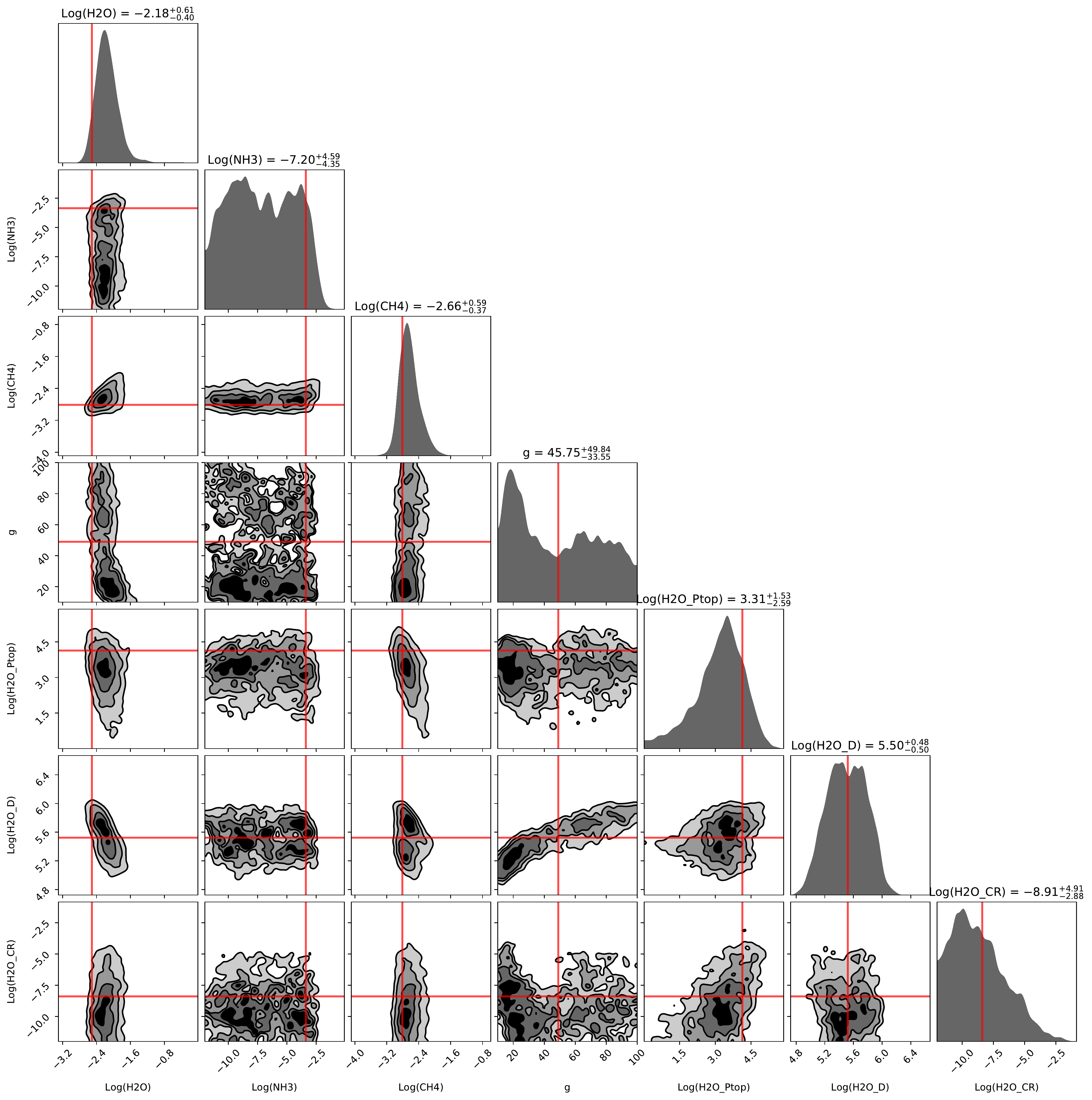}
		\caption{Posterior distribution of the free parameters of the model for the \upsand scenario. The red lines indicate the ground truths of the synthesized model. The numbers reported on top of the 1-D distributions are relative to the median and 1$\sigma$ values of the distributions. The correlation between the CH$_4$ and P$_{top}$ is weak. In this example the correlation between the VMR$_{H_{2}O}$ and both VMR$_{CH_{4}}$ and D$_{cld,H_2O}$ can be seen, however, the relative 2-D distributions are quite localized. No multi-modal solutions have been found. \label{fig:ups_post}}
	\end{figure}
	\pagebreak
	
	\subsection{47 uma b - water cloud}
	\begin{figure}[!h]
		\plotone{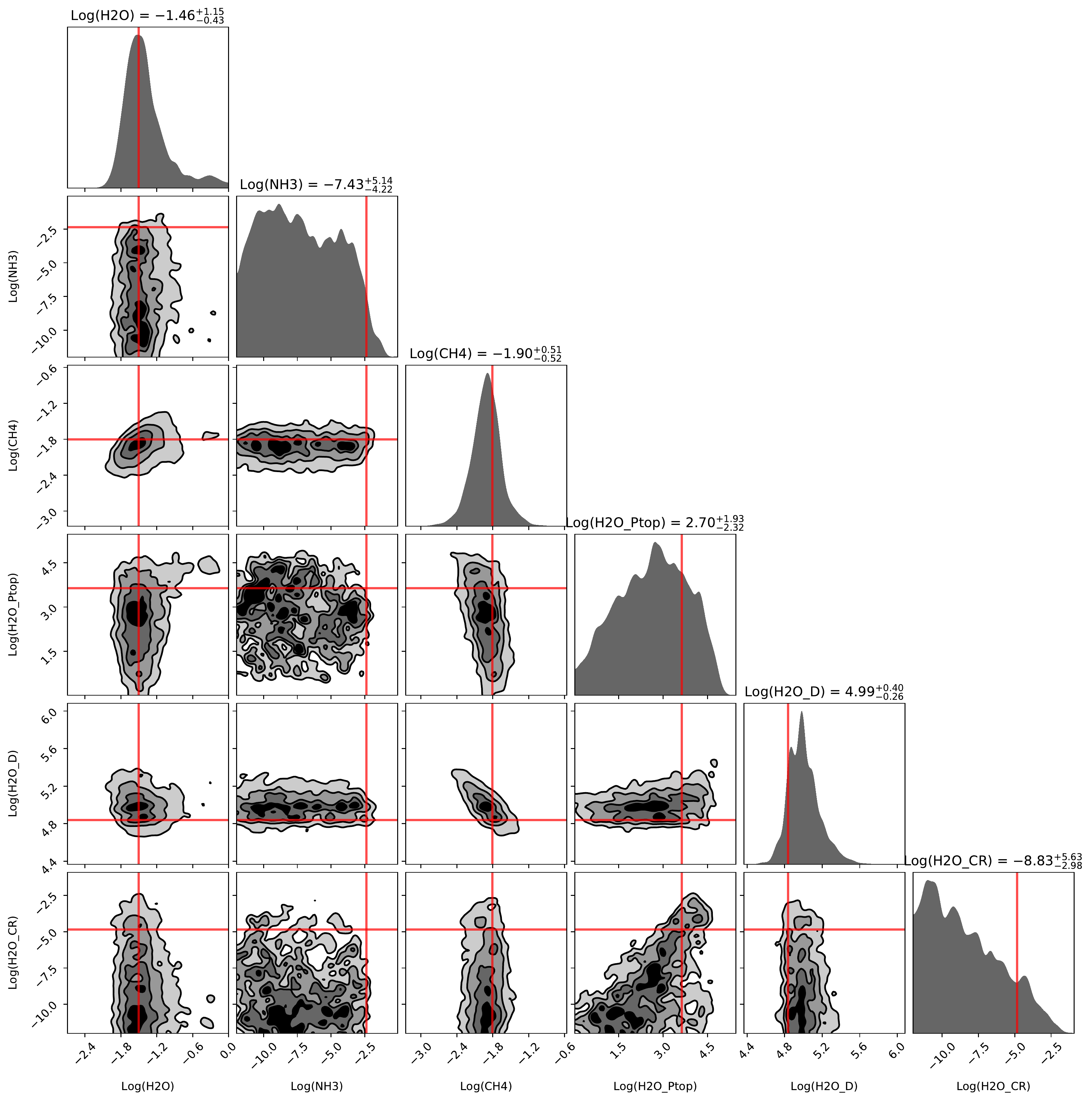}
		\caption{Posterior distribution of the free parameters of the water model for the 47 Uma b science case. The numbers reported on top of the 1-D distributions are relative to the median and 1$\sigma$ values of the distributions. The solid red lines refers to the input parameters (ground truths) used to synthesize the data. \label{fig:uma_post}}
	\end{figure}
	\pagebreak
	
	\subsection{Jupiter ( \Jupiter ) - 2-clouds model}
	\begin{figure}[!h]
		\plotone{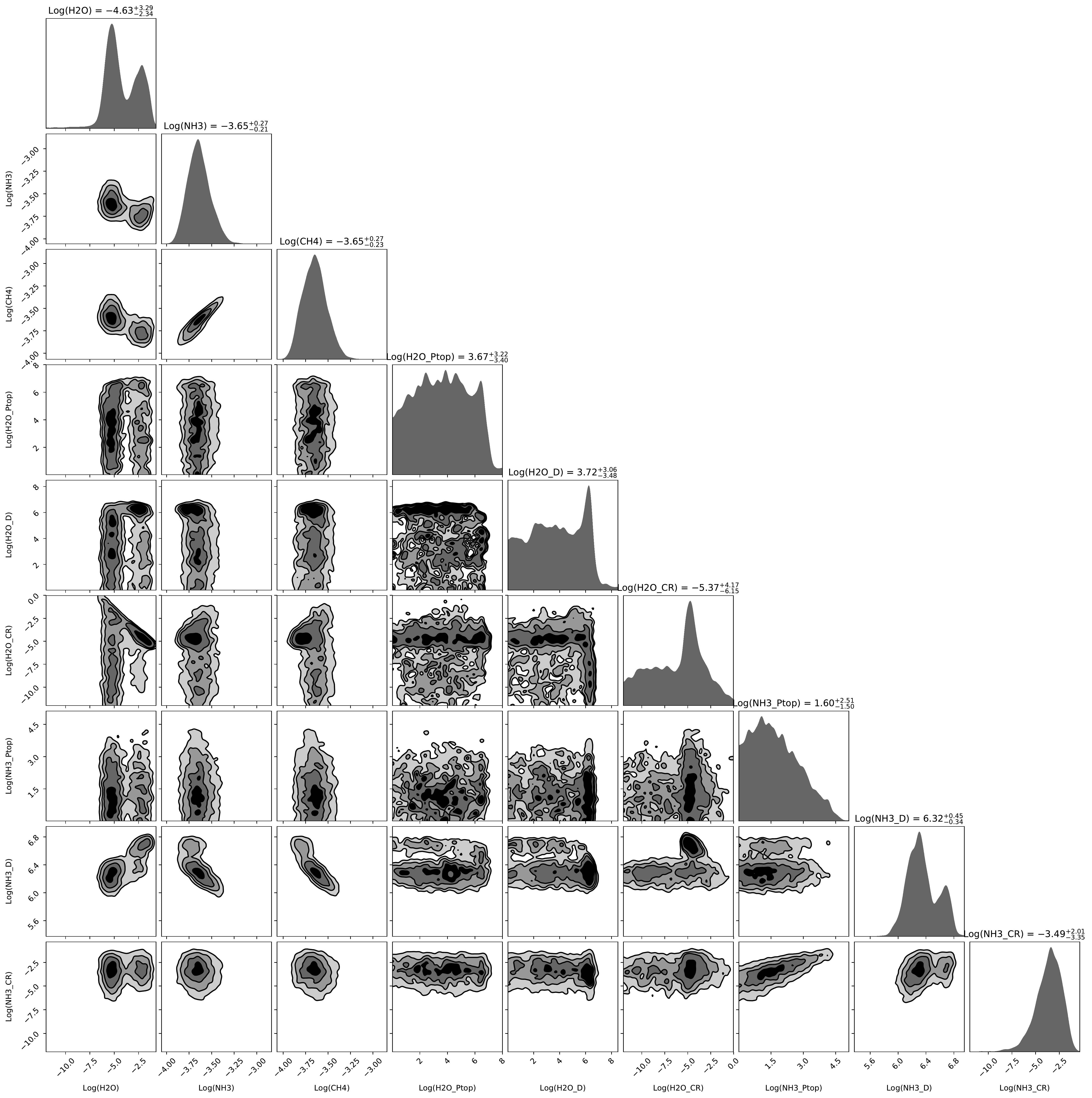}
		\caption{Posterior distribution of the free parameters of the 2-clouds model for the Jupiter scenario. The numbers reported on top of the 1-D distributions are relative to the median and 1$\sigma$ values of the distributions. \label{fig:jup_post}}
	\end{figure}
	\pagebreak
	
	\subsection{Jupiter ( \Jupiter ) - ammonia cloud}
	\begin{figure}[!h]
		\plotone{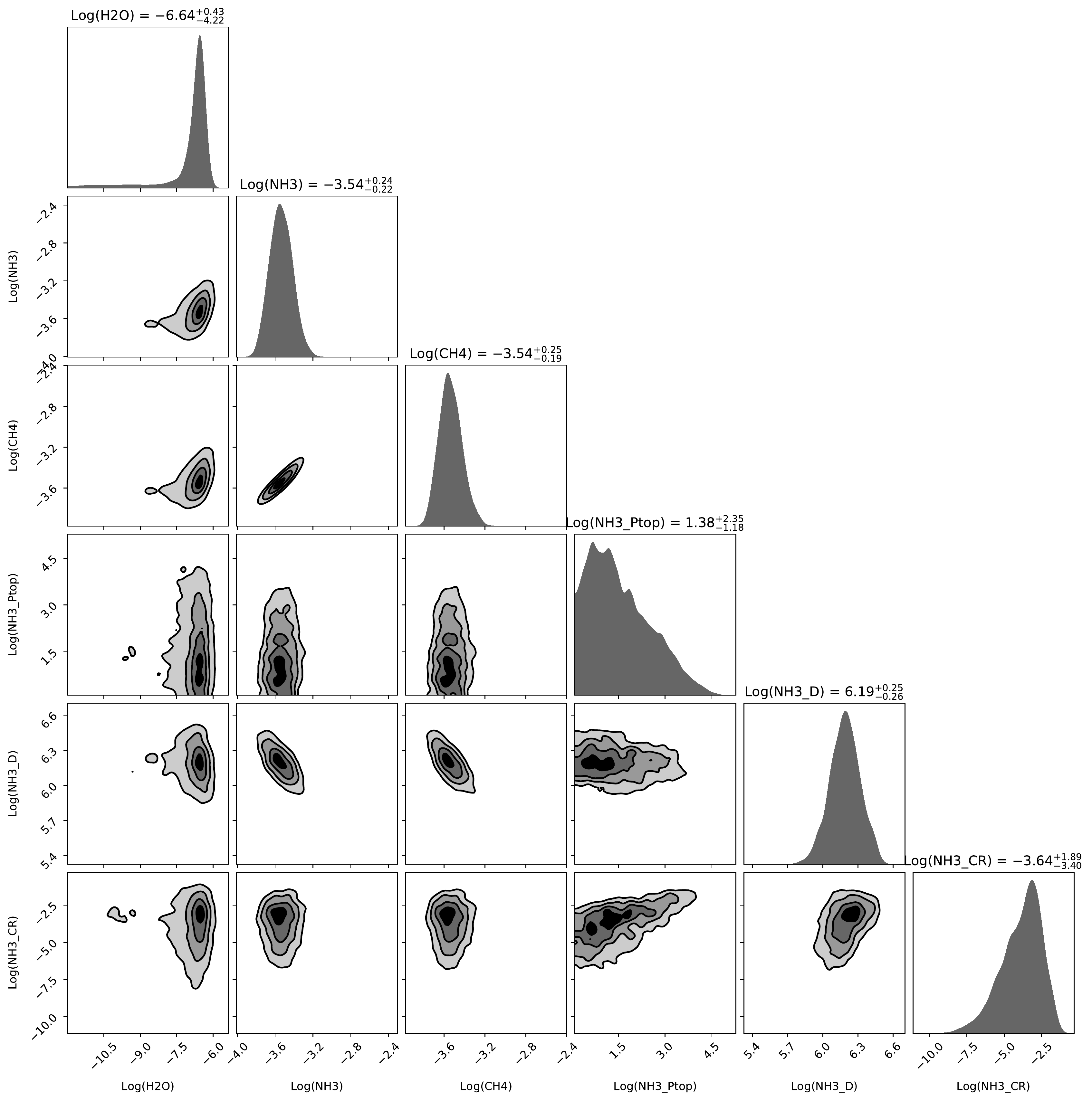}
		\caption{Posterior distribution of the free parameters of the ammonia cloud model for the Jupiter science case. The numbers reported on top of the 1-D distributions are relative to the median and 1$\sigma$ values of the distributions. \label{fig:jup_post2}}
	\end{figure}
	
\end{document}